\documentclass{aastex631}
\usepackage{amsmath,amssymb,lipsum}
\usepackage{framed}
\usepackage{natbib}

\begin{document}

	\setcounter{page}{1}
	
	\pagestyle{plain}

	\setcounter{page}{1}
	
	\pagestyle{plain}
	
	\title{Traces of Quantum Gravity Effects at Late time Cosmological Dynamics via Distance Measures}
	\author{M. Roushan}\email{m.roushan@umz.ac.ir},\author{N. Rashidi}\email{n.rashidi@umz.ac.ir},\author{K. Nozari}\email{knozari@umz.ac.ir(Corresponding Author)}
	\affiliation{Department of Theoretical Physics, Faculty of Science,
		University of Mazandaran,\\
		P. O. Box 47416-95447, Babolsar, IRAN}

	\begin{abstract}
		Inspired by the entropy-area relation of black hole thermodynamics, we study the thermodynamics of cosmological
		apparent horizon in a spatially flat Friedmann-Robertson-Walker (FRW) universe in the framework of an
		Extended Uncertainty Principle (EUP). The adopted EUP naturally admits a minimal measurable momentum
		(equivalently a maximal measurable length), as an infrared cutoff in the theory. We derive the modified
		Friedmann equations in this setup and explore some predictions of these equations for the late time universe
		via distance measures. We show that in this framework it is possible to realize the late time cosmic speed-up
		and transition to the phantom phase of the equation of state parameter of the effective cosmic fluid without
		recourse to any dark energy component or modified gravity. Inspection of various distance measures in this
		framework shows that an EUP with a negative deformation parameter suffices for the interpretation
		of the late time asymptotically de Sitter universe with standard non-relativistic matter.
	\end{abstract}
	
	\section{Introduction}
	
	Physics is the realization of a deep connection
	between specific rules of black hole physics and the ordinary laws of thermodynamics.
	The compatibility of these two physical concepts, like the main pieces
	of a puzzle, is so that when they are considered together, they create a perfect harmony and
	describe the physical properties of the gravitational systems satisfactorily. The discovery of Hawking
	radiation was the turning point in the correlation between these two fundamental laws.
	In this regard, Hawking revealed that the thermal features of a black hole can be characterized
	by thermodynamic relations~\cite{Hawking1975}. In this way, the thermodynamics of black holes are determined
	according to the entropy and temperature of the black hole~\cite{Bek1973,Bek1974,Wald2001}, and the temperature
	of the event horizon of the black hole corresponds to its surface gravity. Also, in the language
	of general relativity, the black hole entropy-area law explains how
	the entropy of the black hole corresponds to its event horizon area.
	Along this path, Jacobson by considering the relationship between entropy and the area of the
	event horizon and additionally by considering the relationship between the temperature of
	the horizon and surface gravity~\cite{Jacobson1995},  was able to show that Einstein's field equations can directly
	follow the Clausius thermodynamic relationship (the famous fundamental thermodynamic relationship between
	heat, entropy, and temperature as $dQ=TdS$).
	
	In recent years, great progress has been made regarding the connection between gravity and
	thermodynamics; among them the seminal works of Padmanabhan~\cite{Pad2005} and Verlinde~\cite{Verlinde2011}
	have been highly regarded in this field. Jacobson's idea is based on the fact that Einstein's
	field equations are the equations of state specific to spacetime from a thermodynamic point of view.
	In other words, while standard thermodynamics deals with the statistical mechanics of particles that include
	atoms and molecules, gravity deals with the statistical mechanics of spacetime atoms~\cite{Padmanabhan2016}.
	
	As a phenomenological approach to quantum gravity proposal, in recent years authors in this field have focused on the generalized uncertainty
	principle (GUP) and the extended uncertainty principle (EUP). The mechanism of these approaches
	is so that Heisenberg's uncertainty relation is modified in order to incorporate gravitational effects
	in, for instance, Heisenberg's thought experiment~\cite{Adler2010}. In this way, some phenomenological aspects of the high
	energy regime, especially the quantum gravity regime, have been studied in recent years~\cite{Kempf95,Hossenfelder2013,Bosso2023}.
	Also the low energy (large distance) quantum gravitational effect encoded in the extended uncertainty principle is considered, for instance, in Refs~\cite{Hinrichsen1996,Mignemi2010}. In the framework of these scenarios, the underlying quantum field theory includes natural ultraviolet cutoff as a
	minimal length (maximal energy/momentum) and a minimal momentum (maximal length) as infrared cutoff.	
	In the presence of a minimal observable length of the order
	of the Planck length, space finds a granular and discrete structure. In this picture, the fundamental
	cell of spacetime with a size of the order of the Planck length acts as a spacetime atom.
	From the Hawking-Beckenstein relation $S=\frac{A}{4\pi l_{pl}^2}$, we can also realize this
	proportionality, in which $4\pi l_{pl}^2$ represents the surface of a sphere equal to the radius
	of the Planck length, and equivalently, the surface of the spacetime atom.
	
	The existence of a minimal measurable length in various approaches to quantum gravity proposal that
	leads to the generalized uncertainty principle, now is a well-known concept; see for instance ~\cite{Kempf95,Roushan2019,RashidiEPL2023}.
	On the other hand, if we consider the background to be curved, the existence of an observable minimal
	momentum for the test particle is inevitable, and this by itself leads to the extended uncertainty
	principle~\cite{Hinrichsen1996,Mirza2009,Mignemi2010,Noz20}. The extended uncertainty relation with a positive deformation parameter leads to a natural infra-red
	cutoff as a minimal measurable momentum or equivalently a maximal measurable length in the gravitational
	structure of the corresponding field theory. In recent years some implications of this infra-red cutoff have been studied in the literature, see for instance~\cite{Filho2016,Chung2019,Dabrowski2020,Lawson2020,Roushan2022,Wagner2022}.
	
	Here in a spatially flat Friedman-Robertson-Walker background, we derive the Friedman equations in the
	presence of an infrared cutoff arising from the extended uncertainty relation. This is done for
	the cosmological apparent horizon inspired by the Bekenstein-Hawking entropy-area relation of the black hole thermodynamics.
	We show that positively accelerated expansion of the late time universe
	can be addressed fully just by the presence of this infra-red cutoff with ordinary, standard matter without the need to recourse to
	unknown dark energy components or modified gravities. We demonstrate that the effective equation of the state parameter of
	the cosmic fluid in the presence of the infra-red cutoff explains the transition to the phantom phase of the universe expansion without the need to introduce unconventional components such as phantom or quintom dark energy candidates. By treating the cosmological distance measures and also statefinder and $O_{m}(z)$ diagnostics in this infrared regularized scenario, we explore the cosmological feasibility of the model. We show that this model is supported confidently by recent observational data.
	
	\section{\label{sec2}Thermodynamics of the Apparent Horizon and Friedmann Equations}
	
	On the basis of the first law of thermodynamics and borrowing the relation between the
	apparent horizon temperature and surface gravity, we derive Friedmann equations for an
	$(n + 1)$-dimensional model universe. For this purpose, the $(n+1)$-dimensional Friedmann
	equation in a thermodynamic perspective can be obtained by taking into account the relation
	$T=\frac{1}{2\pi\tilde{r}_{A}}$ between the temperature and universe's apparent horizon
	$\tilde{r}_{A}$ and the entropy-area relation as $S=\frac{A}{4G}$ (where $G$ and $A$ are the
	Newtonian constant and the apparent horizon area respectively). We consider the line element as
	\begin{equation}
		\label{eq1}
		ds^2=-dt^2+a^{2}(t)\left(\frac{dr^2}{1-k\,r^2}+r^2\,d\Omega_{n-1}^{2}\right)\,,
	\end{equation}
	where $k=0,\pm 1$ indicates the spatial curvature constant and
	$d\Omega_{n-1}^{2}$ reveals the line element of an $(n-1)$-dimensional
	unit sphere. We express the line element (\ref{eq1}) in the following form
	\begin{equation}
		\label{eq2}
		ds^2=h_{ij}\,dx^i\,dx^{j}+\tilde{r}^{2}\,d\Omega_{2}^{2}\,.
	\end{equation}
	where the quantities are defined as $h_{ij}=(-1,\,a^{2}/(1-kr^2))$, $\tilde{r}= a(t)\,r$
	and $x^{i}=(t,r)$. The dynamics of the apparent horizon radius $\tilde{r}_{A}$ is obtained
	using this metric~\cite{Cai05}. This horizon, the outermost of all trapped surfaces, is a
	marginally outer trapped surface with vanishing outer null expansion. It is a sphere of radius
	$\tilde{r}_A$ that fulfills the following requirement~\cite{Cha10,Cha11}
	\begin{eqnarray}
		\label{eq3}
		h^{ij}\,\partial_{i}\tilde{r}\,\partial_{j}\tilde{r}=0\,.
	\end{eqnarray}
	By solving this equation, we can obtain the following relation for the location of the apparent horizon in the FRW flat universe
	\begin{eqnarray}
		\label{eq4} \tilde{r}_{A}=a\,r=\frac{1}{H}\,,
	\end{eqnarray}
	where $H$ is the Hubble parameter, $H=\frac{\dot{a}}{a}$ and an over dot marks a cosmic time derivative.
	Now, we assume a universe consisting of a perfect fluid with energy-momentum tensor as
	\begin{eqnarray}
		\label{eq5} T_{\mu\nu}=(\rho+p)\,u_{\mu}\,u_{\nu}-p\,g_{\mu\nu}\,.
	\end{eqnarray}
	In this relation, $u_{\nu}$ is $4$-velocity, and $\rho$ and $p$ are the energy density
	and pressure of perfect fluid respectively. Application of the conservation law for $T^{\mu\nu}$
	provides the conservation equation as
	\begin{eqnarray}
		\label{eq6} \dot{\rho}+3\,H\,(\rho+p)=0\,.
	\end{eqnarray}
	Since the \textit{Work Function} is crucial in using the first law of thermodynamics,
	we express this function in terms of the trace of the energy-momentum tensor as follows~\cite{Hay98,Bak00}
	\begin{eqnarray}
		\label{eq7} W=-\frac{1}{2}T^{ij}\,g_{ij}\,.
	\end{eqnarray}
	It is noteworthy that $T^{ij}$ is considered as a projection of the ($3+1$)-dimensional
	energy-momentum tensor $T^{\mu\nu}$ in the normal direction of the 2-sphere~\cite{Cai05}.
	By assuming a flat FRW universe filled with perfect fluid, the above equation turns to
	\begin{eqnarray}
		\label{eq8} W=\frac{1}{2}(\rho-p)\,.
	\end{eqnarray}
	We use the first law of thermodynamics expressed as
	\begin{eqnarray}
		\label{eq9} dE=T\,dS+W\,dV\,,
	\end{eqnarray}
	where the parameter $E$ marks the total energy within the apparent horizon which is known also as
	the\emph{ Misner-Sharp energy} and it is determined by
	\begin{eqnarray}
		\label{eq10} E=\rho\,V\,.
	\end{eqnarray}
	Here $V$ is the volume of a $3$-dimensional sphere given as $V=\frac{4}{3}\pi\tilde{r}_{A}^{3}$, where
	$\tilde{r}_{A}$ is the radius of the apparent horizon. By differentiating both sides of the relation (\ref{eq10}), we find
	\begin{eqnarray}
		\label{eq11} dE=\rho\,dV+V\,d\rho\,.
	\end{eqnarray}
	In this step, by using conservation equation (\ref{eq6}) and relations (\ref{eq9}) and (\ref{eq10}) we arrive at the following relation
	\begin{eqnarray}
		\label{eq12}
		dE=4\pi\tilde{r}_{A}^{2}\,\rho\,d\tilde{r}_{A}-4\pi\,\tilde{r}_{A}^{3}(\rho+p)\,H\,dt\,.
	\end{eqnarray}
	To proceed, the second term on the right hand side of the relation (\ref{eq9}) is obtained as the following form
	\begin{eqnarray}
		\label{eq13}
		W\,dV=2\pi\,\tilde{r}_{A}^{2}\,(\rho-p)\,d\tilde{r}_{A}\,.
	\end{eqnarray}
	Our next step is to obtain the first term of equation (\ref{eq9}), $T\,dS$ versus $\tilde{r}_{A}$ to fix the second term. For this reason,
	we apply the relation between temperature and surface gravity as~\cite{Awa14}
	\begin{eqnarray}
		\label{eq14} T=\frac{\kappa}{2\pi}\,.
	\end{eqnarray}
	where the surface gravity $\kappa$ is given as
	\begin{eqnarray}
		\label{eq15}
		\kappa=\frac{1}{2\sqrt{-h}}\,\partial_{i}\bigg(\sqrt{-h}\,h^{ij}\,\partial_{j}\tilde{r}_{A}\bigg)\,.
	\end{eqnarray}
	Using the line element (\ref{eq2}), $\kappa$ finds the following form
	\begin{eqnarray}
		\label{eq16}
		\kappa=-\frac{1}{\tilde{r}_{A}}\bigg(1-\frac{\dot{\tilde{r}}_{A}}{2\,H\,\tilde{r}_{A}}\bigg)\,.
	\end{eqnarray}
	Generally, the entropy-area relation has the following form
	\begin{eqnarray}
		\label{eq17} S=\frac{f(A)}{4G}\,.
	\end{eqnarray}
	Therefore, the quantity $T\,dS$ can be calculated from equations (\ref{eq14}), (\ref{eq15}), and (\ref{eq17}) as follows
	\begin{eqnarray}
		\label{eq18}
		T\,dS=-\frac{1}{2\,\pi\,\tilde{r}_{A}}\Bigg(1-\frac{\dot{\tilde{r}}_{A}}{2\,H\,\tilde{r}_{A}}\Bigg)d\Bigg(\frac{f(A)}{4\,G}\Bigg)
		=-\frac{1}{2\,\pi\,\tilde{r}_{A}}\Bigg(1-\frac{\dot{\tilde{r}}_{A}}{2\,H\,\tilde{r}_{A}}\Bigg)\Bigg(\frac{d\,f(A)}{dA}\Bigg)
		\Bigg(\frac{8\,\pi\,\tilde{r}_{A}}{4\,G}\,d\tilde{r}_{A}\Bigg)\,.
	\end{eqnarray}
	Using equation (\ref{eq4}), its time derivative and also equations (\ref{eq9}), (\ref{eq12}),(\ref{eq13}) and (\ref{eq18}),
	we determine the second Friedmann equation as follows
	\begin{eqnarray}
		\label{eq19} \dot{H}\,\frac{d\,f(A)}{dA}=-4\pi\,G\,(\rho+p)\,.
	\end{eqnarray}
	By considering the conservation equation (\ref{eq6}) and integrating equation (\ref{eq19}) we derive
	\begin{eqnarray}
		\label{eq20} \rho=-\frac{3}{2\,G}\int \frac{f'(A)}{A^2}\,dA\,,
	\end{eqnarray}
	where $f'(A)=\frac{d\,f(A)}{dA}$. As usual, for $f(A)=A$ the standard Friedmann
	equations are recovered
	\begin{eqnarray}
		\label{eq21} \dot{H}=-4\pi\,G\,(\rho+p)\,,
	\end{eqnarray}
	\begin{eqnarray}
		\label{eq22} H^{2}=\frac{8\pi\,G}{3}\rho\,.
	\end{eqnarray}
	After a brief preliminaries and introduction of notations and conventions,
	now we are going to see the effects of an extended uncertainty relation with negative deformation parameter in a cosmological
	setup. This EUP comes into play via the entropy-area relation, resulting in the late time modified Friedmann equations.

	\section{\label{sec3}Modified Friedmann equations in the extended uncertainty principle framework}
	
	The uncertainty principle in the presence of minimal momentum as a natural infrared cutoff is modified as follows~\cite{Mignemi2010}
	\begin{eqnarray}
		\label{eq23} \Delta x\Delta p \geq
		\hbar\Big(1+\eta(\Delta x)^2\Big)\,.
	\end{eqnarray}
	This equation implies the existence of a minimal measurable momentum for negative $\eta$ as $\Delta p_{min}=\hbar\sqrt{-\eta}=\hbar\sqrt{|\eta|}$.
This can be shown as follows:

From the relation (\ref{eq23}), the minimum value of $\Delta x \Delta p$  can be written as
\begin{eqnarray}
		\label{eq23a} \Delta x\Delta p =
		\hbar\Big(1+\eta(\Delta x)^2\Big)\,.
	\end{eqnarray}
On the other hand, we have the lowest value of $\Delta x$ on the right hand side of the Eq.~(\ref{eq23a}) from the \emph{standard uncertainty relation} as follows
$$\Delta x \Delta p\geq \hbar \rightarrow \Delta x \geq\frac{\hbar}{\Delta p}\rightarrow (\Delta x)_{_{min}}|_{\eta=0}=\frac{\hbar}{\Delta p}\,.$$
Therefore, relation (\ref{eq23}) turns into the following relation in leading order
\begin{eqnarray}
		\label{eq23b} \Delta x\Delta p \geq
		\hbar\Big(1+\eta\frac{\hbar^{2}}{(\Delta p)^{2}}\Big)\,.
	\end{eqnarray}

Since essentially $\Delta x \Delta p \geq 0$; it is clear that the above relation would be meaningless if its right hand side to be negative. So, we set the right hand side to be greater than or equal to zero. That is,
$$\hbar\Big(1+\eta\frac{\hbar^{2}}{(\Delta p)^{2}}\Big)\geq 0\,.$$
which naturally results in
$$(\Delta p)_{min}\simeq \hbar\sqrt{-\eta}=\hbar\sqrt{|\eta|}\,,$$
since in our case $\eta$ is negative, $\eta<0$.  Therefore, for negative deformation parameter $\eta$, we have a \emph{real} $(\Delta p)_{min}$, and our model rely on the physical existence of a \emph{real} minimal measurable momentum. We note that the authors in Ref.~\cite{Du2022} have similarly applied this technique to obtain the minimal length in their framework (see the equations (36) to (38) of this reference). Also, the authors in Ref.~\cite{Luo2023} have studied a novel higher-order extended uncertainty principle (EUP) that maintains a minimum length $(\Delta x)_{min}\simeq \hbar\sqrt{|\eta|}$.

Now, with the EUP as the relation (\ref{eq23}), we get $\Delta x$ as follows
	\begin{eqnarray}
		\label{eq24} \Delta x \gtrsim
		\frac{1}{2\hbar\eta}\Bigg(\Delta p-\sqrt{(\Delta
			p)^{2}-4\,\hbar^2\,\eta}\Bigg)\,.
	\end{eqnarray}
	
This solution is only half of the correct formula since, as a quadratic in $\Delta x$, the EUP (\ref{eq23}) for positive $\eta$ implies that
	$$\frac{1}{2\hbar\eta}\Bigg(\Delta p-\sqrt{(\Delta
		p)^{2}-4\,\hbar^2\,\eta}\Bigg)\leq \Delta x \leq \frac{1}{2\hbar\eta}\Bigg(\Delta p+\sqrt{(\Delta
		p)^{2}-4\,\hbar^2\,\eta}\Bigg).$$ For $\eta < 0$, the domains of validity are disconnected. The existence of a minimal measurable momentum as an infrared cutoff could be essentially a model-independent effect, much similar to the existence of a minimal measurable length (as an ultraviolet cutoff) that is addressed in all approaches to quantum gravity proposal~\cite{Calmet2004}. It has been shown in the literature that the deformation parameters in the context of GUP and EUP models could be positive or negative in essence. For instance, we address some studies in this field as follows: the authors in Ref.~\cite{Jizba2010} have studied GUP on the lattice and they have deduced negative GUP parameter. Subsequently, Scardigli and Casadio have investigated the Hawking temperature in the presence of GUP effects through a Wick rotation of an effective Schwarzschild-like metric using a negative GUP parameter~\cite{Scardigli2015}. The authors in Ref.~\cite{Carr2015} have obtained the features of sub-Planckian black holes with negative GUP parameters. Ong in Ref.~\cite{Ong2018} has claimed that the Chandrasekhar limit is not established using positive values of the GUP parameter, while the negative GUP parameter explains this inconsistency. The thermodynamic evolution and phase transition of a black hole in the presence of quantum gravity effects with positive and negative GUP parameters have been analyzed in Ref.~\cite{Zhou2022}. Also, in Ref.~\cite{Bouninfante2019}, the authors have studied some concepts of the Generalized Uncertainty Principle (GUP) in the framework of black hole physics. In this regard, they have discussed the sign of the deformation parameter of the theory and concluded that when a corpuscular quantum description of gravitational interaction is adopted, there is a possibility of this parameter being negative.
Scardigli in Ref.~\cite{Scard2019} pointed out that a negative deformation parameter indicates the existence of a sharp, classical world at the Planck Scale.
As a more precise explanation, the most probable trajectories in the phase spaces of our universe as a quantum system are classical trajectories. In this way, by assuming negative values for the deformation parameter, we can also have classical systems.
Also, the authors in Ref.~\cite{Luo2023}, investigated a new higher order EUP within the context of gravitational baryogenesis to explain the baryon asymmetry of the universe (BAU). They have estimated the bounds of the EUP parameters for both positive and negative cases.  	
	Actually the sign of the deformation term(s) has its origin in Modified Dispersion Relations (MDRs)~\cite{Amelino2006}. As has been stated in references~\cite{Park2008,Bolen2005} and also  ~\cite{Ghosh2024} (with a typo in this later reference), the positive $\eta$ implies an AdS (Anti de Sitter) space, whereas the negative $\eta$ can correspond to a dS (de Sitter) space with lower bound of $(\Delta p)_{min}=\hbar\sqrt{-\eta}=\hbar\sqrt{|\eta|}\simeq\hbar\sqrt{\Lambda}$, where $\Lambda$ is a positive cosmological constant (see alse~\cite{Gine2020} and ~\cite{Ong2020}). The current cosmological observations suggest that we live at late time in an asymptotically de Sitter universe. The EUP with negative parameter, $\eta < 0$, in our model gives rise to a minimal momentum of the mentioned order with identification $\sqrt{-\eta}\simeq\sqrt{\Lambda}$, that is, $\sqrt{|\eta|}\simeq\sqrt{\Lambda}$ corresponding to a de Sitter phase of cosmic expansion. It is worth mentioning that a positive $\eta$ in Eq.~(\ref{eq23}) has not the capability to result in an observationally viable cosmology.

	The alteration in the surface area of a black hole that absorbs or radiates a
	quantum particle is provided by $\Delta A\geq 8\pi\l_{pl}^{2}\,E\,R$, where $R$ and $E$
	are respectively the physical size of the particle and the energy of the particle~\cite{Chr70}.
	By consideration of the existence of a finite limit for the particle energy, we have
	$\Delta A_{min}\geq 8\pi\,l_{pl}^{2}\,c\,\Delta p\,\Delta x$. Therefore, using Eq.~(\ref{eq20}), we find
	\begin{eqnarray}
		\label{eq25} \Delta A_{min} \gtrsim \frac{4\pi\,l_{pl}^2}{\hbar\,\eta}\Delta p\Bigg(\Delta p-\sqrt{(\Delta
			p)^{2}-4\hbar^2\,\eta}\Bigg)\,.
	\end{eqnarray}
	Now we suppose $(\Delta x)^{2}=\frac{A}{\pi}$, so that $(\Delta p)^{2}=\frac{\pi}{A}\hbar^2$.
	Therefore, the alteration in the area is given by~\cite{Awa14}
	\begin{eqnarray}
		\label{eq26} \Delta A_{min} \simeq \chi\,
		\frac{4\pi^2\,l_{pl}^2}{\eta\,A}\Bigg(1-\sqrt{1-\frac{4A}{\pi}\eta}\Bigg)\,.
	\end{eqnarray}
	where we have set $\hbar=1=c$. We note that at this stage we have used the standard Heisenberg uncertainty as $\Delta x \Delta p\geq \hbar$ since adopting EUP Eq. (\ref{eq23}) in this case generates higher order products of uncertainties that can be neglected for these essentially infinitesimal quantities. To be more specific, we should replace $\Delta x$ with $\Delta x(1+\eta(\Delta x)^2)$. Then $(\Delta x)^2$ transforms to $(\Delta x)^2+2 \eta(\Delta x)^4+\eta^2 (\Delta x)^6 $ which clearly shows that the second and third terms can be neglected easily. It is worth mentioning that the role of the deformation parameter, $\eta$, has not been excluded in our calculations since it comes into play from Eq. (\ref{eq24}) and is seen clearly in Eq. (\ref{eq25}). We just neglected higher order products of uncertainties.
	
The parameter $\chi$ can be determined via the Bekenstein-Hawking entropy-area relation.
By adopting a calibration as $\Delta S_{min}=b=\ln (2)$~\cite{Ada04} and having the
Bekenstein-Hawking entropy formula, we take $\frac{b}{\chi}\equiv2\,\pi$~\cite{Med04}. Therefore,
	\begin{eqnarray}
		\label{eq27}\frac{dS}{dA}=\frac{\Delta S_{min}}{\Delta A_{min}}
		=\frac{\eta\,A}{2\pi\,l_{pl}^2\Bigg(1-\sqrt{1-\frac{4A}{\pi}\eta}\Bigg)}\,.
	\end{eqnarray}
	Using equation ~(\ref{eq17}) to find $f'(A)$, and then equations (\ref{eq20}) and (\ref{eq27})
	along with the relation $A=4\pi\,\tilde{r}_{A}^{2}$, we derive the energy density as follows
	\begin{eqnarray}
		\label{eq28}\rho=\frac{3H\sqrt{H^2-16\eta}}{16\pi l_{pl}^2}-\frac{3\eta}{\pi l_{pl}^2}\ln\Big(H+\sqrt{H^2-16\eta}\Big)
		+\frac{3H^2}{16\pi l_{pl}^2},
	\end{eqnarray}
	With some manipulations in equation (\ref{eq28}), we derive
	\begin{eqnarray}
		\label{eq29}\frac{G}{2l_{pl}^2}H^2+\frac{GH\sqrt{H^2-16\eta}}{2\,l_{pl}^2}-\frac{8G\eta}{l_{pl}^2}
		\ln\Bigg(H+\sqrt{H^2-16\eta}\Bigg)
		=\frac{8\pi\,G}{3}\,\rho\,,
	\end{eqnarray}
	In relation (\ref{eq29}), we separate the correction terms that appear due
	to the quantum gravitational effects of EUP and the standard term as follows
	\begin{eqnarray}
		\label{eq30}H^2=\frac{8\pi\,G}{3}\,\rho+\frac{8\pi\,G}{3}\,\rho_{_{EUP}}\,.
	\end{eqnarray}
	where by definition,
	\begin{eqnarray}
		\label{eq31}\rho_{_{EUP}}\equiv\frac{3}{8\pi\,G}\Bigg(\bigg(1-\frac{G}{2l_{pl}^2}\bigg)H^2-\frac{G}{2l_{pl}^2}H\sqrt{H^2
			-16\eta}+\frac{8\,G\eta}{l_{pl}^2}\ln\bigg(H+\sqrt{H^2-16\eta}\bigg)\Bigg)\,.
	\end{eqnarray}
	Therefore, we derive Modified Friedmann equations by using Eqs. (\ref{eq19}), (\ref{eq27}) and (\ref{eq29}) as
	
	\begin{eqnarray}
		\label{eq32}\frac{G}{2l_{pl}^2}H^2+\frac{G}{2\,l_{pl}^2}H\sqrt{H^2-16\eta}-\frac{8G\eta}{l_{pl}^2}
		\ln\Bigg(H+\sqrt{H^2-16\eta}\Bigg)
		=\frac{8\pi\,G}{3}\,\rho\,,
	\end{eqnarray}
	
	\begin{eqnarray}
		\label{eq33}\dot{H}\Bigg(\frac{8\,G\,\eta}{H^2\,l_{pl}^2}\Bigg)\Bigg(\frac{1}{1-\sqrt{1-\frac{16\eta}{H^2}}}\Bigg)=-4\pi\,G\,(\rho+p)\,.
	\end{eqnarray}
	After the derivation of all required EUP-corrected equations, now we study late time cosmological dynamics in this setup.
	
	\section{\label{sec4}Some cosmological parameters in the context of EUP}
	In this section, we study the effect of the extended uncertainty principle on some important cosmological parameters.
	
	\subsection{Cosmic scale factor}
	In the standard model of cosmology, the expansion of the universe is measured by a single
	function of time called the scale factor, $a(t)$. In fact, all cosmological
	analysis and treatments on the cosmic evolution of cosmological observables are parametrized by the
	scale factor. In this paper, our aim is to obtain the corrected scale factor taking into account
	the EUP (\ref{eq23}) with negative deformation parameter, $\eta$. For this purpose, solving the Friedmann
	equation (\ref{eq32}) and by considering $\rho=\rho_0\,{a^{-3}(t)}$ and definition of $H=\frac{\dot{a}(t)}{a(t)}$,
	we get the corrected cosmic scale factor as follows
	\begin{eqnarray}
		\label{eq34}a(t)=\left(\frac{t}{4\pi\,l_{pl}^{2}\,\rho_0+\sqrt{8\pi\,l_{pl}^{2}\,\rho_0\left(2\pi \,l_{pl}^{2}\,\rho_0-3\eta\,t^{2}\right)}}\right)^{l_{pl}{\sqrt{\frac{8\pi\,\rho_0}{3}}}}\,.
	\end{eqnarray}
	
	Figure 1 shows the second order derivative of the scale factor in the presence of quantum gravitational
	effects as the EUP (\ref{eq23}) with negative deformation parameter. From this figure, one can explicitly see that
	the universe experiences a thermal accelerated expansion at late time. In fact, this result
	can be thought-provoking considering its compatibility with observational data today. Note that in
	our setup, this result is obtained just by considering the standard matter but by taking into account
	the EUP (\ref{eq23}) with negative deformation parameter.
	\begin{figure}
		\begin{center}\includegraphics{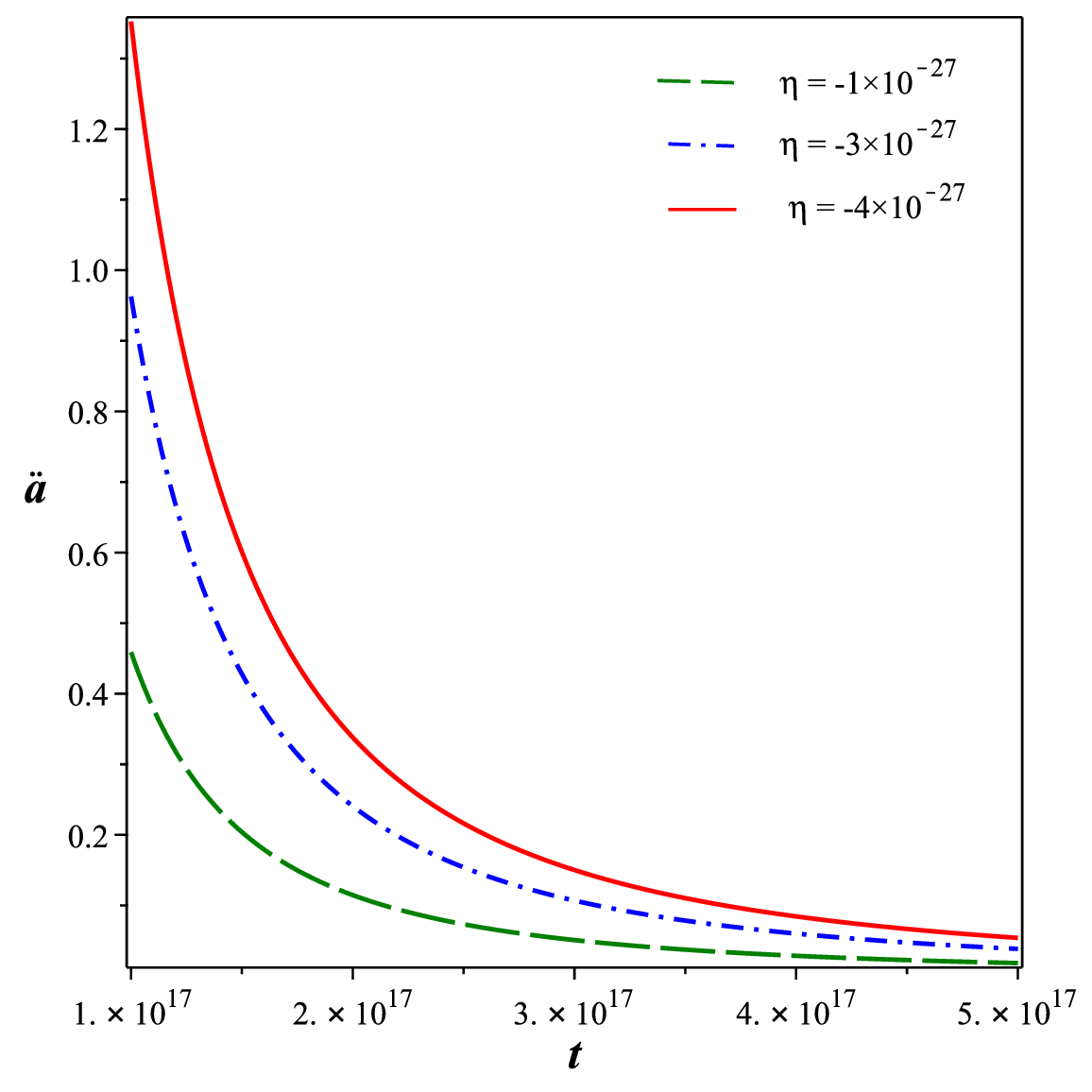}\vspace{6.5cm}
		\end{center}
		\caption{\label{fig1}\small {Evolution of the second order derivative of the scale factor, $\ddot{a}$, as a
				function of cosmic time in the case of the Extended Uncertainty Principle as (\ref{eq23}) with a negative deformation parameter. As is seen, the universe is positively accelerated with this EUP in the presence of ordinary matter at late time.}}
	\end{figure}
	
	\subsection{Deceleration Parameter}
	
	To specify the acceleration in an expanding universe, the historical deceleration parameter
	plays a crucial role. In such a way that a negative value of the deceleration parameter, that is, $q < 0$, demonstrates cosmic
	acceleration of the model under investigation, while positive $q > 0$ represents a decelerating universe.
	The deceleration parameter can be determined as
	\begin{eqnarray}
		\label{eq35}q=-1-\frac{\dot{H}}{H^2}\,.
	\end{eqnarray}
	In the case of EUP, by using the formulation of $\dot{H}$ from Eq. (\ref{eq33}), we get
	\begin{eqnarray}
		\label{eq36}\dot{H}=-\frac{4\pi\rho_0\,l_{pl}^{2}\Biggl[{\cal{W}}\!\Big(-\frac{1}{16\eta}\exp\{{\frac{-2\pi\,l_{pl}^{2}\rho_0\left(1+z\right)^{3}
					+3\eta}{3\eta}}\}\Big)-1\Biggl]}{{\cal{W}}\!\left(-\frac{1}{16\eta}\exp\{{\frac{-2\pi\,l_{pl}^{2}\rho_0(1+z^3)+3\eta}{3\eta}}\}\right)}\left(1+z\right)^{3}\,,
	\end{eqnarray}
	where, ${\cal{W}}\!$ is the \textit{Lambert Function}. Then by inserting equation (\ref{eq34}) into equation (\ref{eq33}), we obtain the following
	expression for the deceleration parameter
	\begin{eqnarray}
		\label{eq37}q=\frac{\Biggl[-{\cal{W}}\!\Big(-\frac{1}{16\eta}\exp\{{\frac{-2\pi\,l_{pl}^{2}\rho_0\left(1+z\right)^{3}
					 +3\eta}{3\eta}}\}\Big)+1\Biggl]\eta-\pi\,l_{pl}^{2}\rho_0\left(1+z\right)^{3}}{\Biggl[{\cal{W}}\!\Big(-\frac{1}{16\eta}\exp\{{\frac{-2\pi\,l_{pl}^{2}\rho_0\left(1+z\right)^{3}
					+3\eta}{3\eta}}\}\Big)-1\Biggl]\eta}\,.
	\end{eqnarray}
	In Figures 2, 3, and 4 we plot the variation of the deceleration parameter with redshift for the late time
	universe using three sources of observational data.
	From the plots, we see that the deceleration parameter is negative in late time (which has been
	identified by a dashed line), so exhibiting the late time cosmic positive acceleration.
	Once again, we note that this cosmic speed up is a result of the EUP (\ref{eq23}) with negative deformation parameter in the presence
	of just ordinary non-relativistic matter.

	\begin{figure}
		\begin{center}\includegraphics{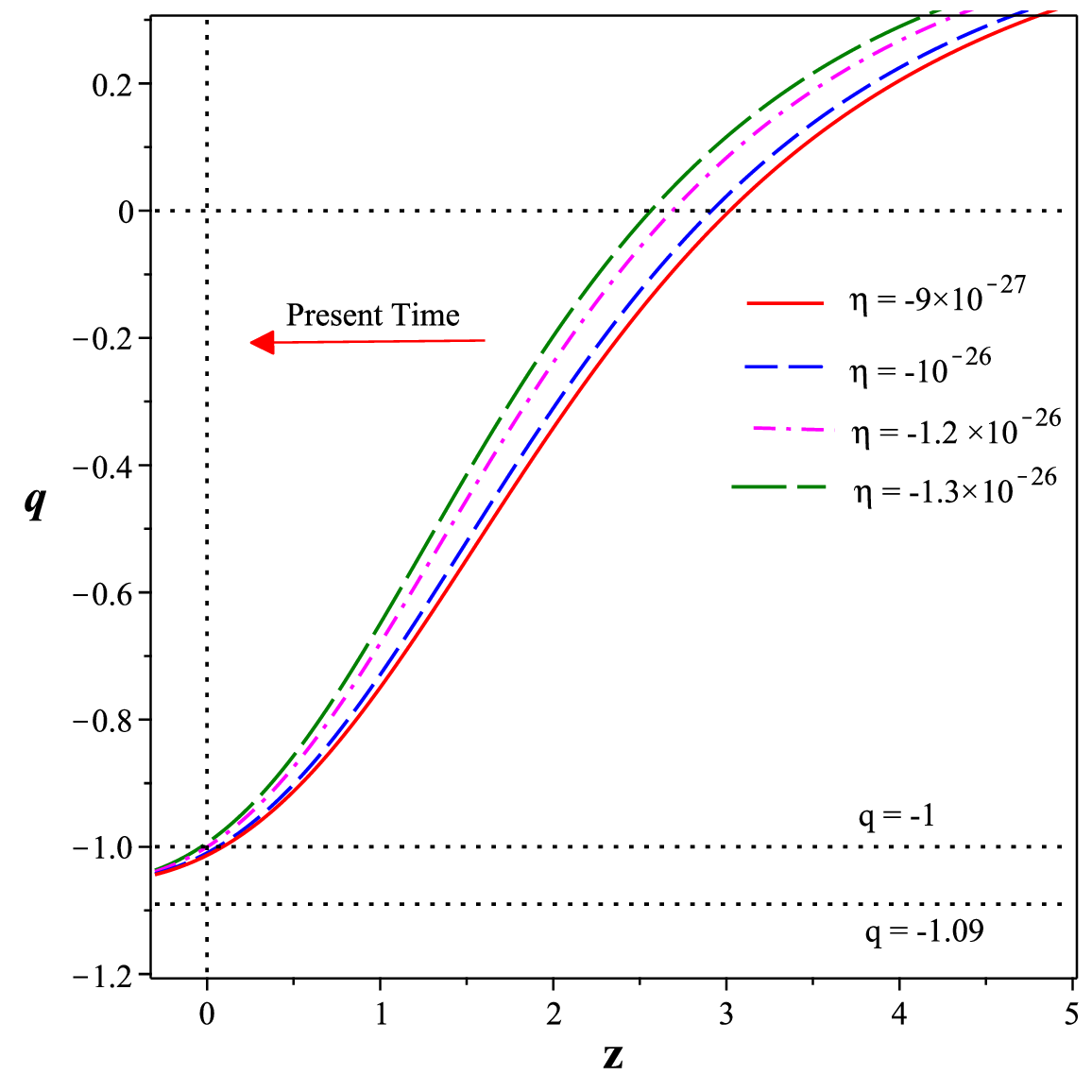}\includegraphics{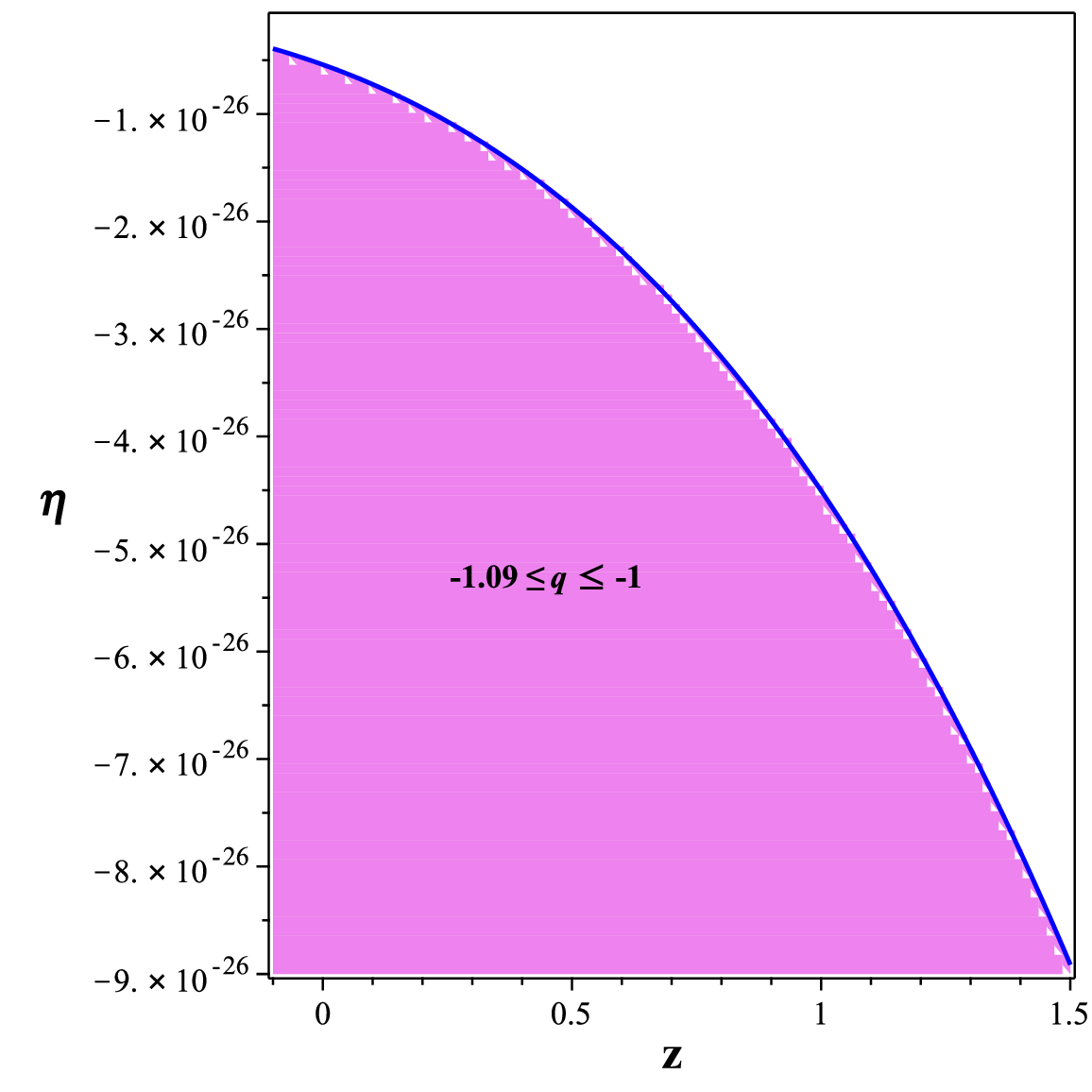}\vspace{6.5cm}
		\end{center}
		\caption{\label{fig2}\small {Evolution of the deceleration parameter $q$ as a
				function of redshift (left panel) using Planck2018 data~\cite{Pl18a}.
				The right panel shows the allowed values of $\eta$ to have late-time cosmic speed up in this setup.}}
	\end{figure}
	
	\begin{figure}
		\begin{center}\includegraphics{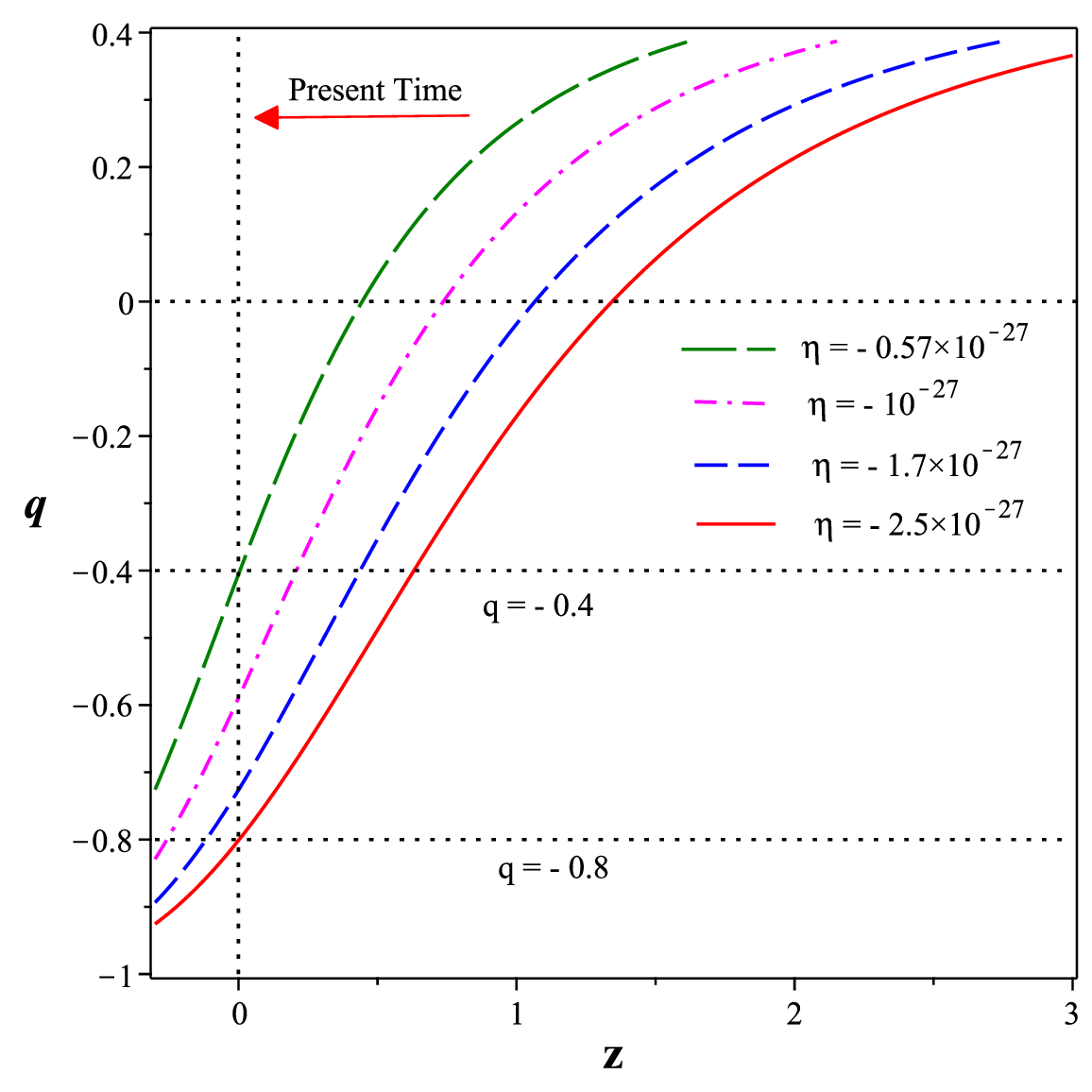}\includegraphics{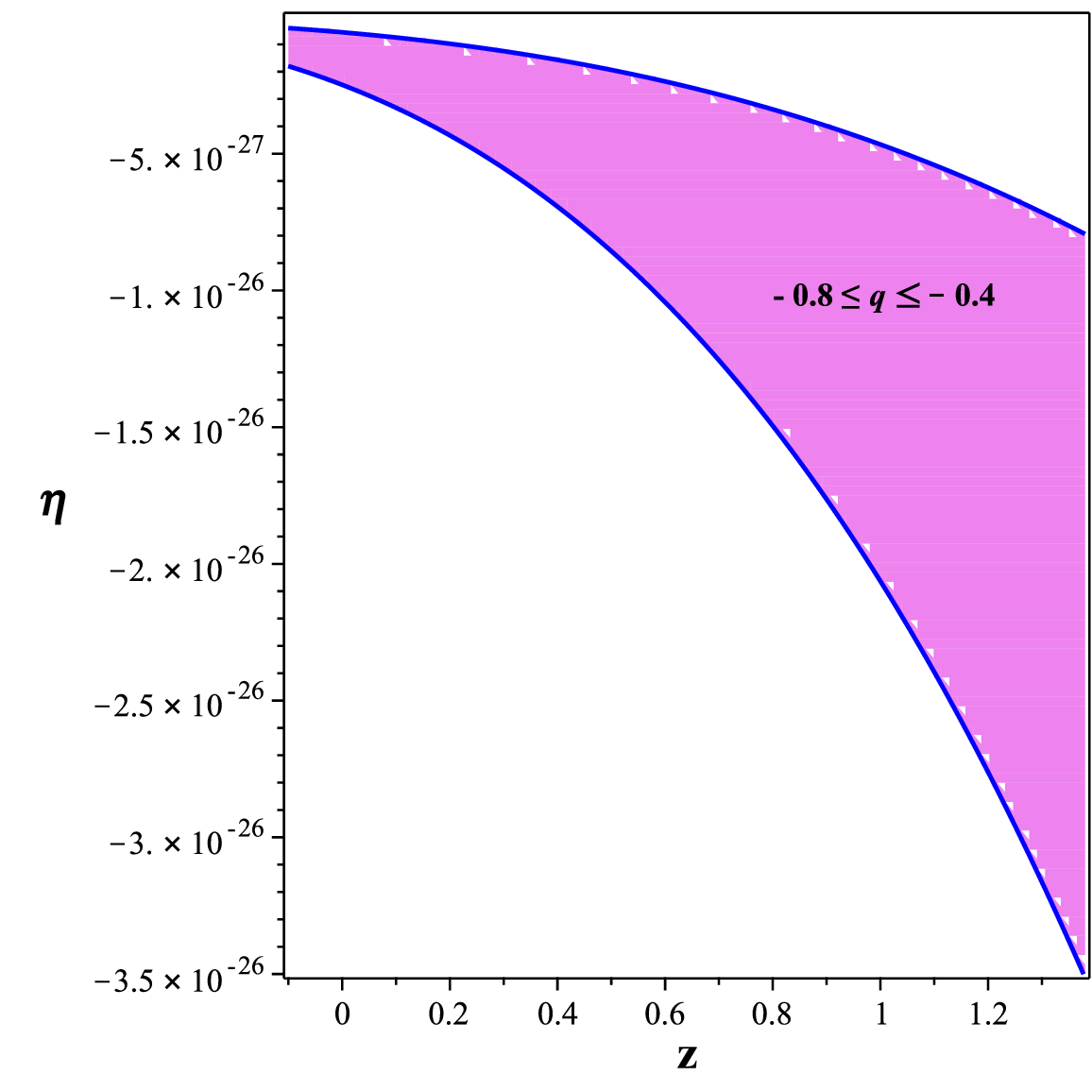}\vspace{6.5cm}
		\end{center}
		\caption{\label{fig3}\small {Evolution of the deceleration parameter $q$ as a
				function of redshift (left panel) using the SNIa data~\cite{Mamon2017,Avesta2022}.
				The right panel shows the allowed values of $\eta$ to have late-time cosmic speed up in this setup.}}
	\end{figure}
	
	\begin{figure}
		\begin{center}\includegraphics{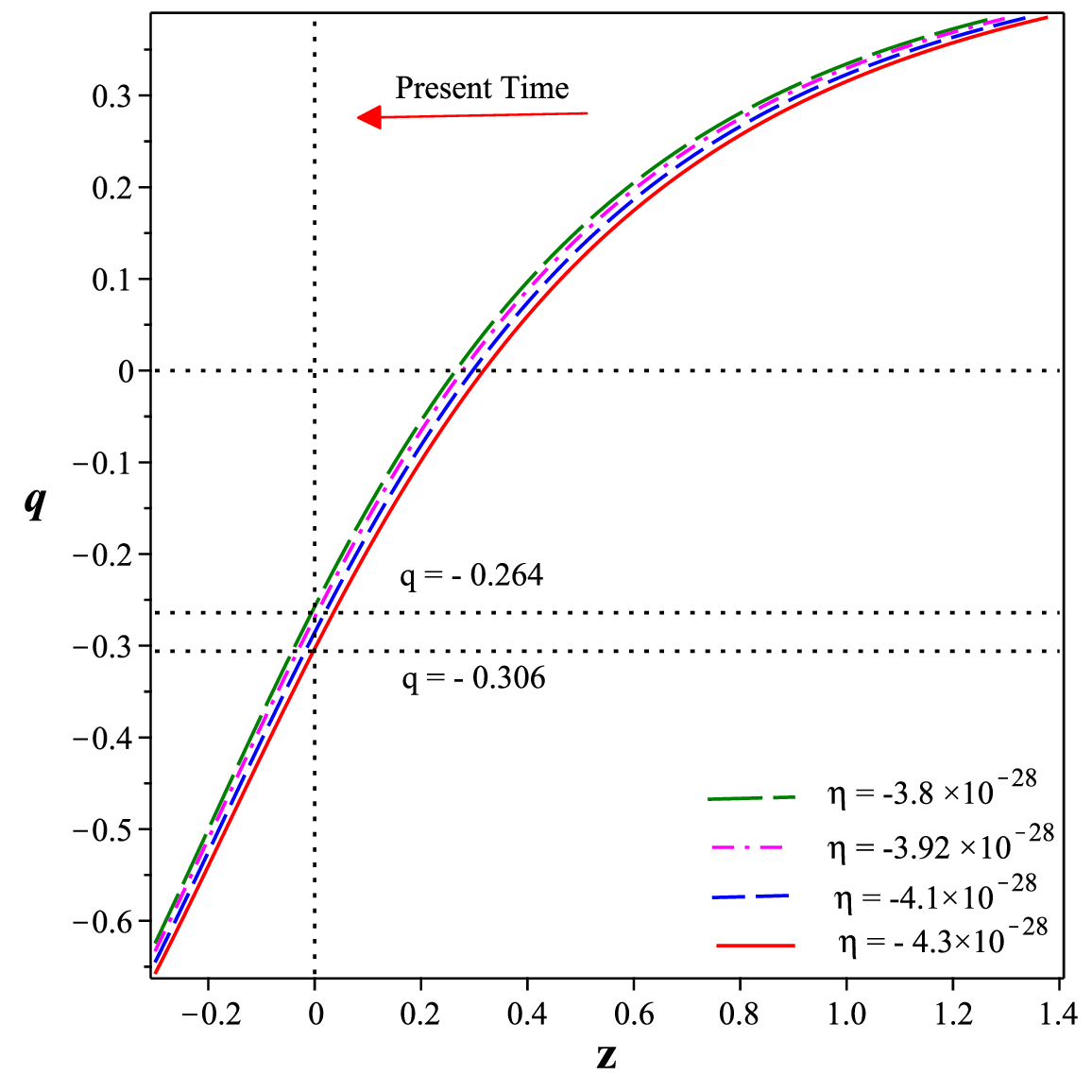}\includegraphics{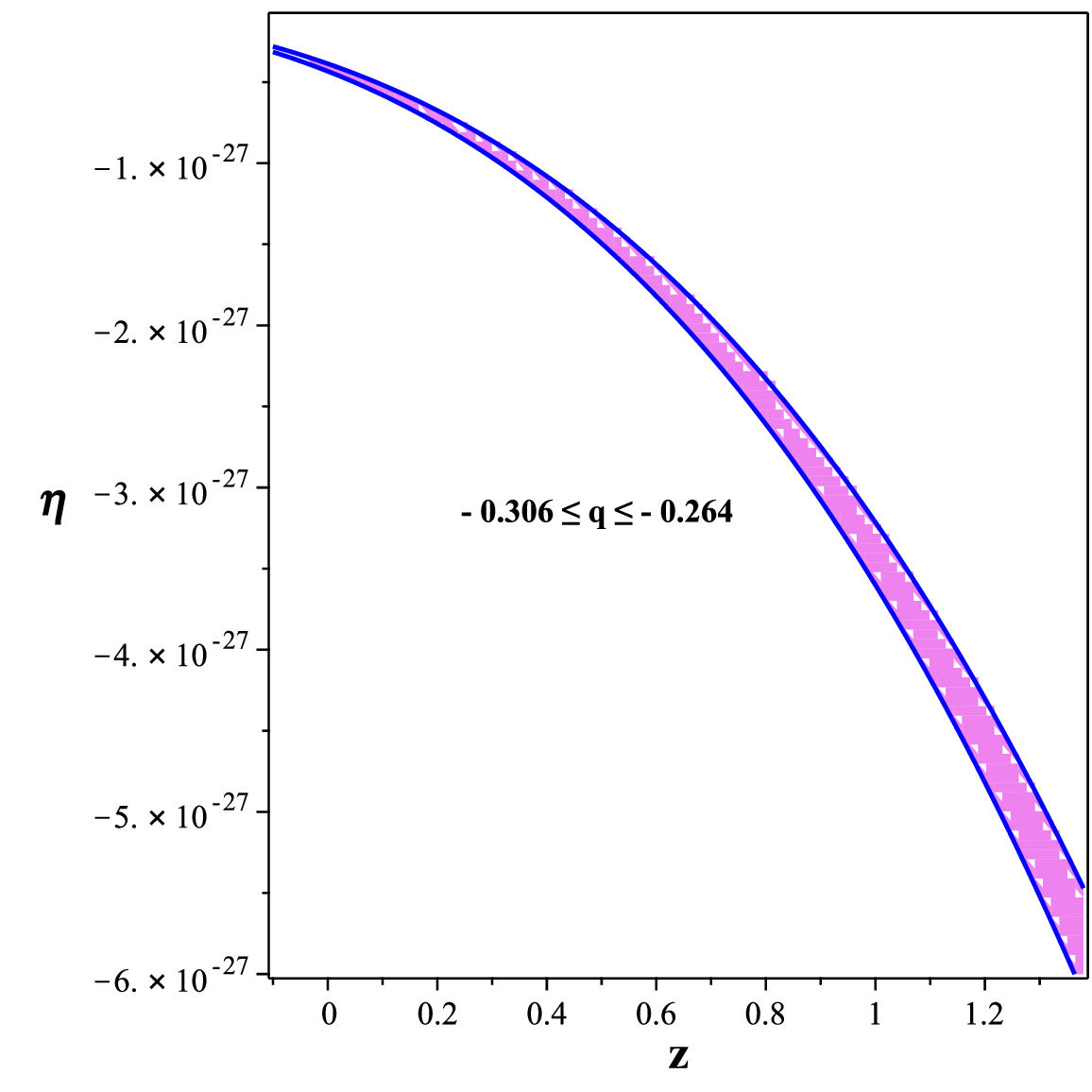}\vspace{6.5cm}
		\end{center}
		\caption{\label{fig4}\small {Evolution of the deceleration parameter $q$ as a
				function of redshift (left panel) using Hubble+Pantheon+BAO data $q_{0}=-0.285\pm 0.021$~\cite{Myrza2023}.
				The right panel shows the allowed values of $\eta$ to have late-time cosmic speed up.}}
	\end{figure}
	
	\section{\label{sec5}Cosmic effective equation of state parameter and the model observational viability}
	Equation of state parameter is generally intended for classifying cosmological models.
	For the universe to be positively accelerated driven by a quintessence field,
	the equation of state parameter is restricted to $-1<w<-\frac{1}{3}$.
	For the cosmological constant, the equation of state parameter is given by $w=-1$.
	The universe is currently in a phantom phase of expansion, encoded by $w<-1$~\cite{Pl18a}.
	In the present study, we show that just by considering the non-relativistic matter but
	in the context of the EUP (\ref{eq23}) with negative deformation parameter, the effective equation of state parameter of the cosmic fluid
	evolves from the quintessence phase toward the phantom phase in a satisfactorily cosmic evolution, crossing
	the cosmological constant (phantom divide) line, $w=-1$, which is consistent with the recent observational data.
	For the purpose of obtaining the EoS parameter, we use the following expression
	\begin{eqnarray}
		\label{eq38}w=\frac{2}{3q}-1\,.
	\end{eqnarray}
	By substituting values from Eq. (\ref{eq37}) into Eq. (\ref{eq38}), we obtain the following relation for the equation of state parameter
	\begin{eqnarray}
		\label{eq39}w=-\frac{2\pi\,l^{2}\rho_0\left(1+z\right)^{3}
			+3\eta\Bigg[{\cal{W}}\!\left(-\frac{1}{16\eta}\exp\{{\frac{-2\pi\,l_{pl}^{2}\rho_0\left(1+z\right)^{3}+3\eta}{3 \eta}}\}\right)-1\Bigg]}{3\Bigg[{\cal{W}}\!\left(-\frac{1}{16\eta}\exp\{{\frac{-2\pi\,l_{pl}^{2}\rho_0\left(1+z\right)^{3}
					+3\eta}{3 \eta}}\}\right)-1\Bigg]\eta}
		\,.
	\end{eqnarray}
	Figure 5 provides the plot of EoS parameter $w$ in this setup as a function of redshift.
	We decode the plot as follows: The EoS parameter crosses the $w=-1$ line at some $z$ in the past for some $\eta$ values
	of the order of $10^{-26}\,-\,10^{-27}$, showing that EUP supports the transition to the phantom phase as required.
	Also, it acts correctly as quintessence in the range $-1<w<-\frac{1}{3}$. Since, observational data
	of Planck2018 which gives $-1.06\leq w\leq-1$ for effective cosmic fluid EoS parameter, we see that
	a model universe containing ordinary matter in the context of Extended Uncertainty Principle with negative deformation parameter is consistent with recent observations.

	\begin{figure}
		\begin{center}\includegraphics{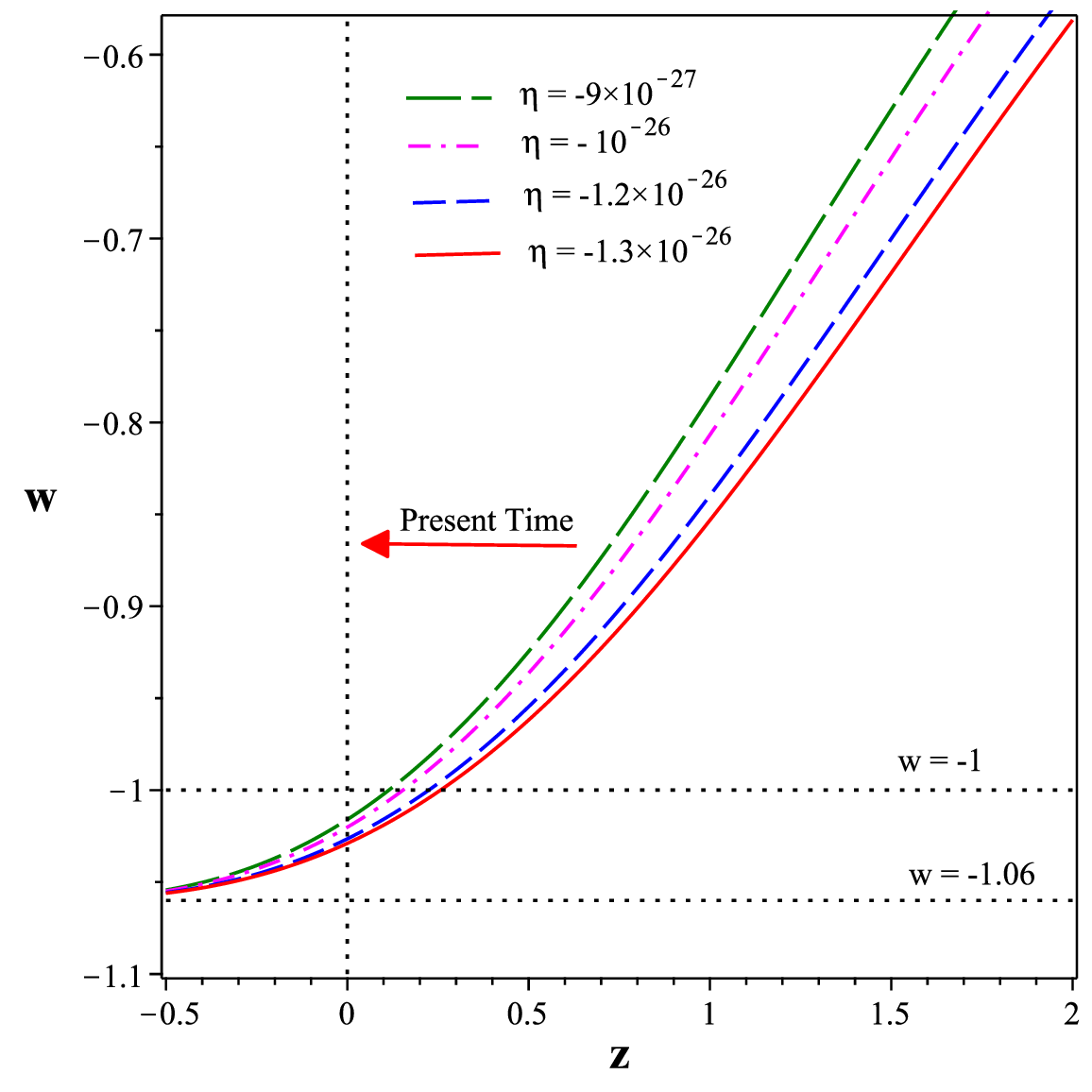}\includegraphics{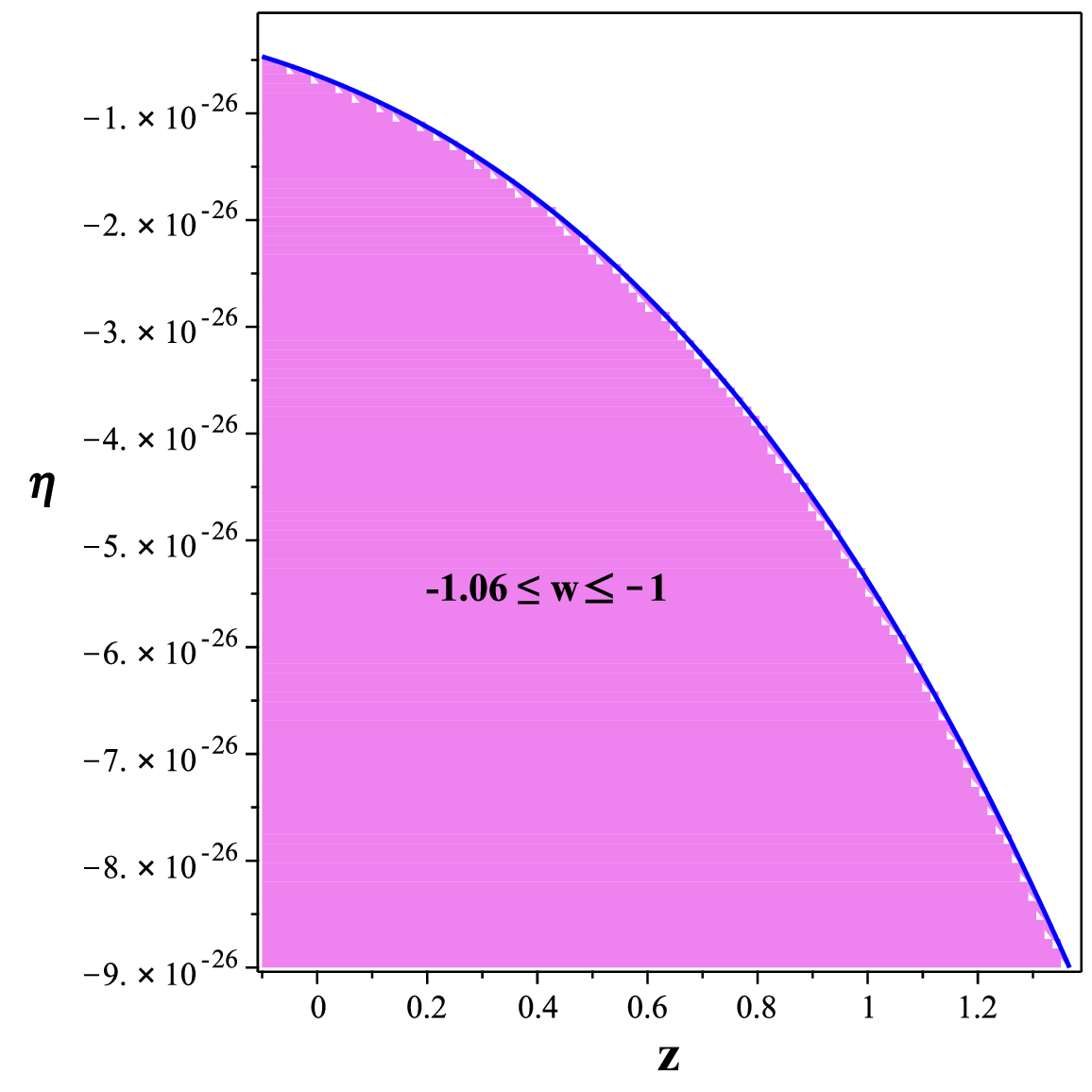}\vspace{6.5cm}
		\end{center}
		\caption{\label{fig5}\small {Cosmic equation of state parameter $w$ as a
				function of cosmic time in the case of Extended Uncertainty Principle (left panel).
				The right panel shows the allowed values of $\eta$ to have late time cosmic speed up.}}
	\end{figure}
	\section{\label{sec6}Statefinder}
	The authors in Ref.~\cite{star2003-1} and ~\cite{star2003-2} have introduced a new mathematical diagnostic
	pair $\{r,s\}$, known as statefinder parameters, which are constructed from the scale factor in order to get a strong
	analysis to divide among various Dark Energy models. The statefinder pair $\{r,s\}$ is specified as
	\begin{eqnarray}
		\label{eq40}r=\frac{\dddot{a}}{aH^3}\,,
	\end{eqnarray}
	
	\begin{eqnarray}
		\label{eq41}s=\frac{r-1}{3(q-\frac{1}{2})}\,.
	\end{eqnarray}
	Therefore, by inserting equation (\ref{eq37}) into (\ref{eq40}) and (\ref{eq41}), we get $r$ versus $s$ as
	\begin{eqnarray}
		\label{eq42}
		r(s)=1-\frac{9}{2}s+\frac{3\pi\,l_{pl}^{2}\rho_0\left(1+z\right)^{3}}
		{\left({\cal{W}}\!\left(-\frac{1}{16\eta}\exp{\{\frac{-2\pi\,l_{pl}^{2}\rho_0\left(1+z\right)^{3}
					+3\eta}{3\eta}\}}\right)-1\right)\eta}s\,.
	\end{eqnarray}
	In Fig. 6, we plot the statefinder diagrams in the $r-s$, $r-q$, and $s-q$ plane for different values of $\eta$ and we compare them with
	$\Lambda$CDM model. As we see the $\Lambda$CDM is in the domain of this cosmological model with IR correction. \\
	\begin{figure}
		\begin{center}\includegraphics{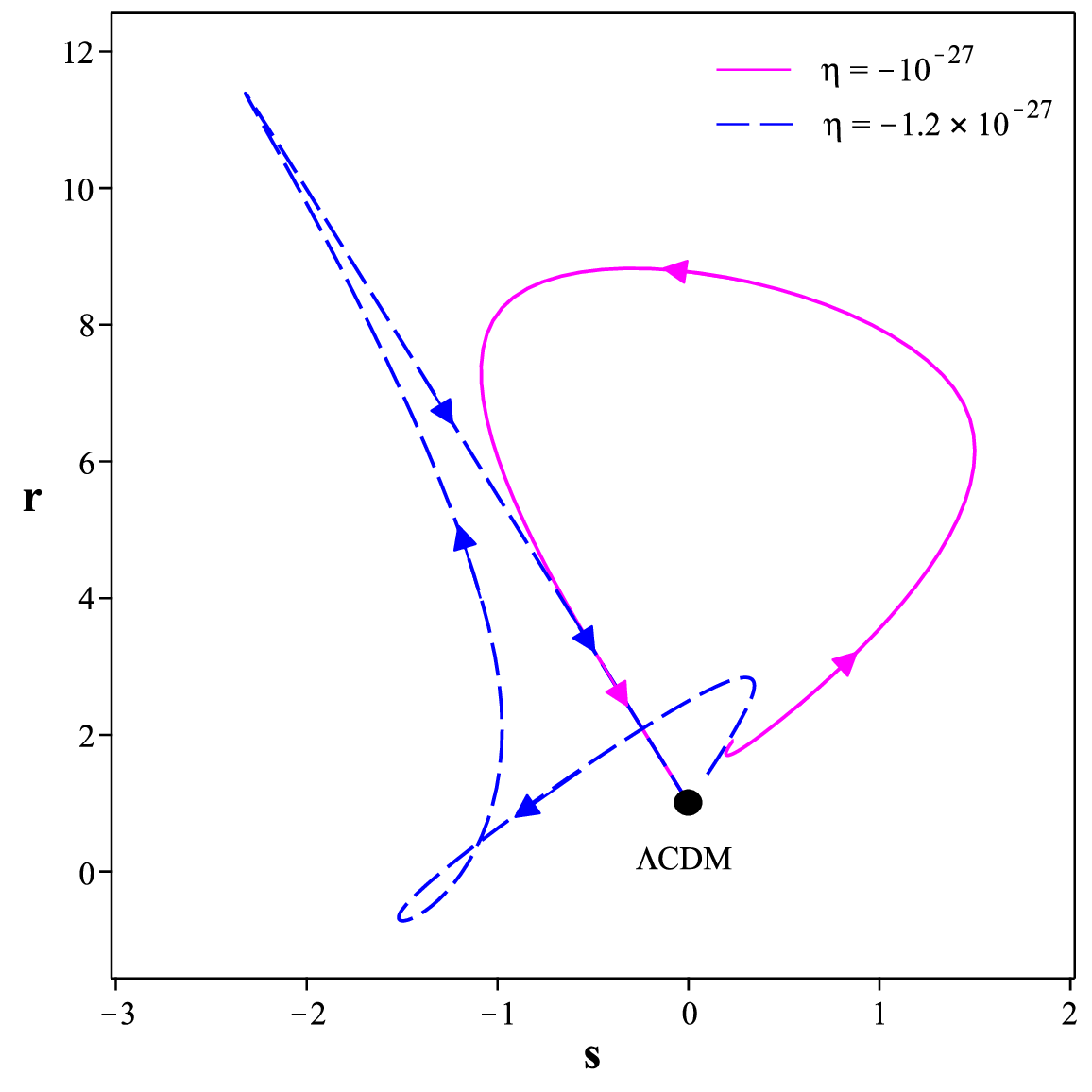}\includegraphics{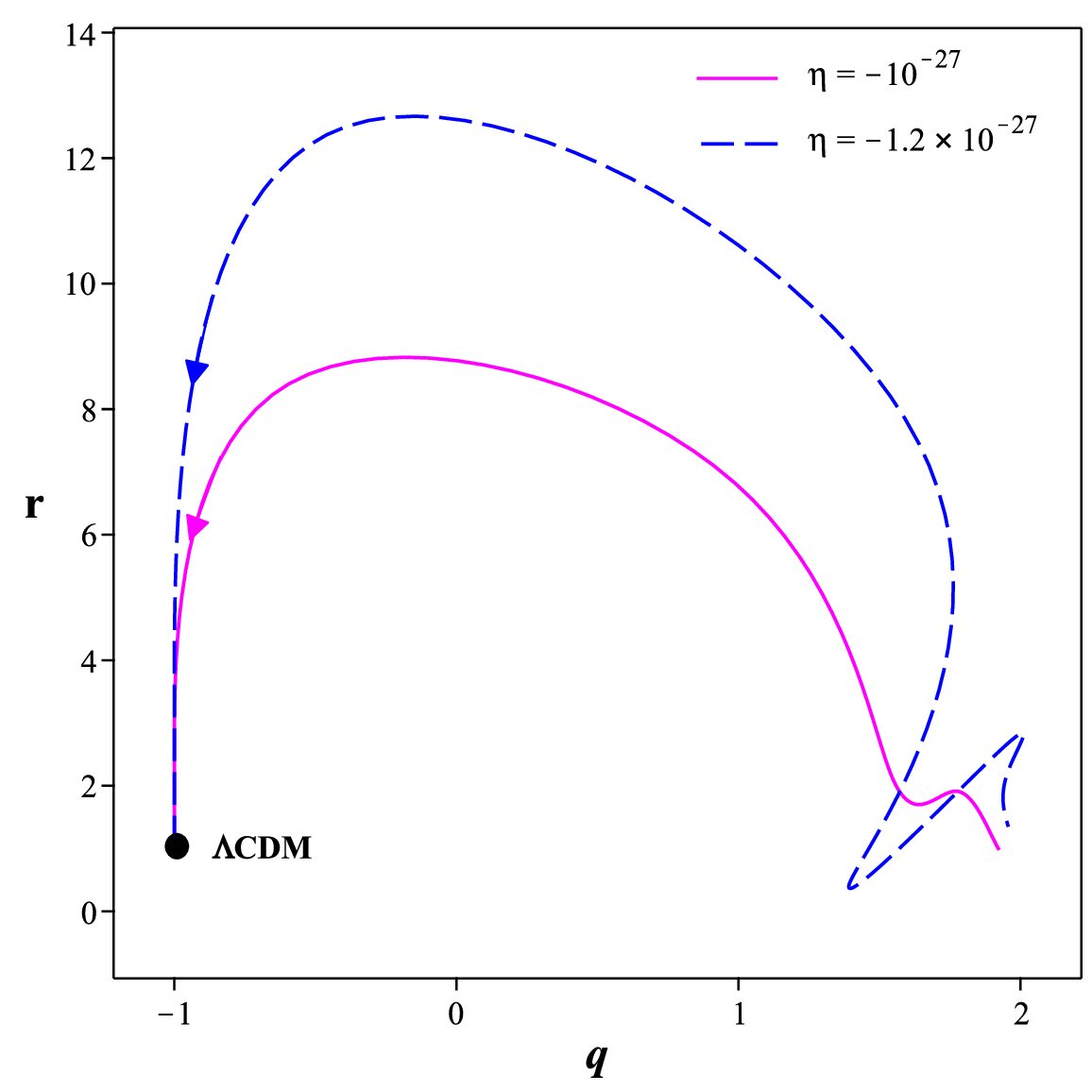}\includegraphics{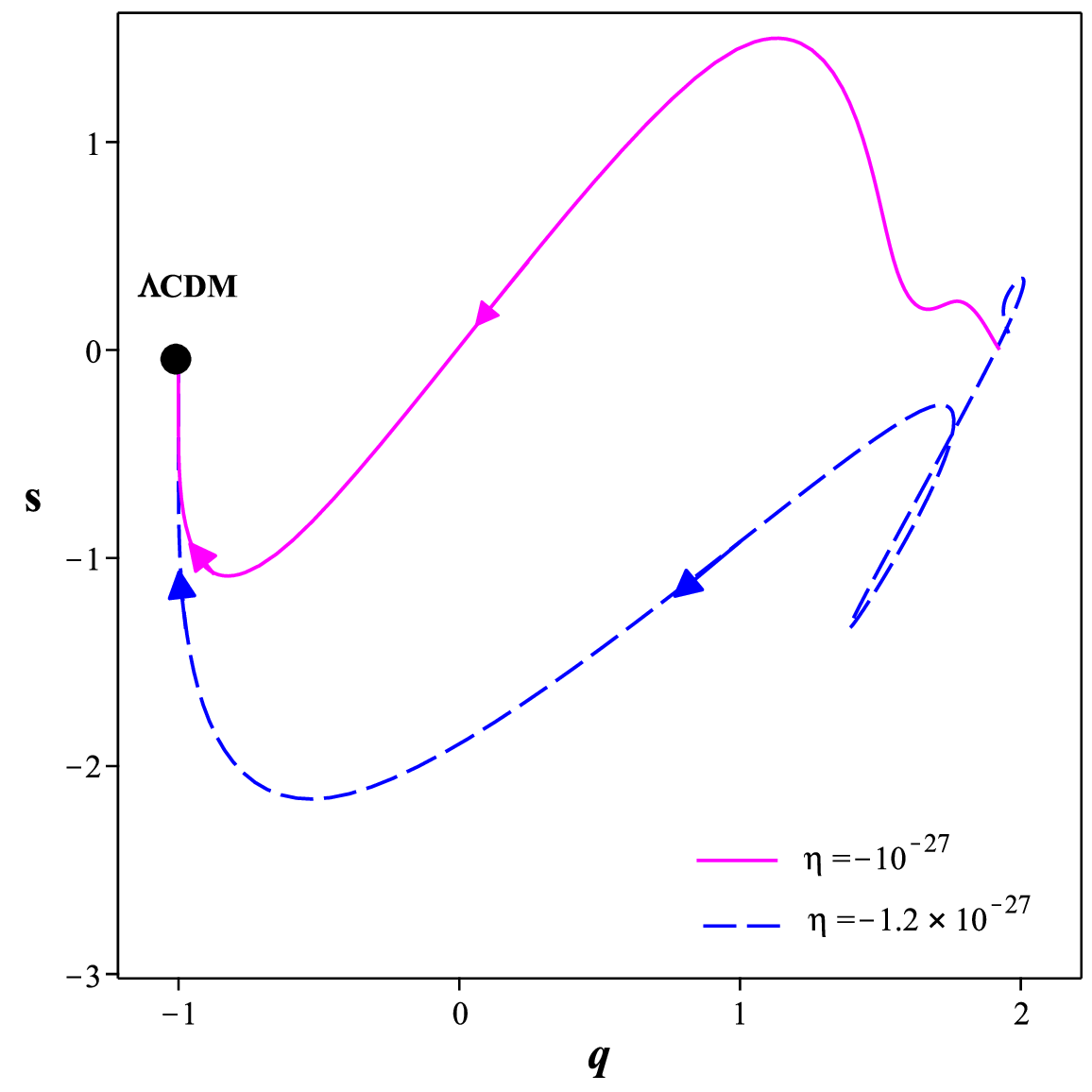}\vspace{6.5cm}
		\end{center}
		\caption{\label{fig6}\small {The behavior of $r-s$, $r-q$, and $s-q$ plane for the constrained values of $\eta$.
				$\Lambda$CDM is in the domain of solutions.}}
	\end{figure}
	Also, the relation between the statefinder parameters $r$ and $s$ in terms of redshift parameter in the context of EUP with negative deformation parameter can be expressed as follows
	\begin{eqnarray}
		\label{eq43}s(z)=-\frac{2\pi\,l_{pl}^{2}\rho_0\left(1+z\right)^{3}}{6\pi\,l_{pl}^{2}\rho_0\left(z +1\right)^{3}
			+9{\cal{W}}\!\left(-\frac{1}{16\eta}\exp{\{\frac{-2\pi\,l_{pl}^{2}\rho_0\left(1+z\right)^{3}
					+3\eta}{3\eta}\}}\right)\eta-9\eta}\,.
	\end{eqnarray}
	
	\begin{eqnarray}
		\label{eq44}r(z)=\frac{\left({\cal{W}}\!\left(-\frac{1}{16\eta}\exp{\{\frac{-2\pi\,l_{pl}^{2}\rho_0\left(1+z\right)^{3}
					+3\eta}{3\eta}\}}\right)-1\right)\eta +\pi\,l_{pl}^{2}\rho_0\left(1+z\right)^{3}}{\left({\cal{W}}\!\left(-\frac{1}{16\eta}\exp{\{\frac{-2\pi\,l_{pl}^{2}\rho_0\left(1+z\right)^{3}
					+3\eta}{3\eta}\}}\right)-1\right)\eta}\,.
	\end{eqnarray}
	Evolution of $r$ and $s$ with redshift for late time universe in the context of EUP with negative deformation parameter is presented in Fig. 7. It is seen that $\Lambda$CDM is in the acceptable domain of this EUP-modified model as required.
	\begin{figure}
		\begin{center}\includegraphics{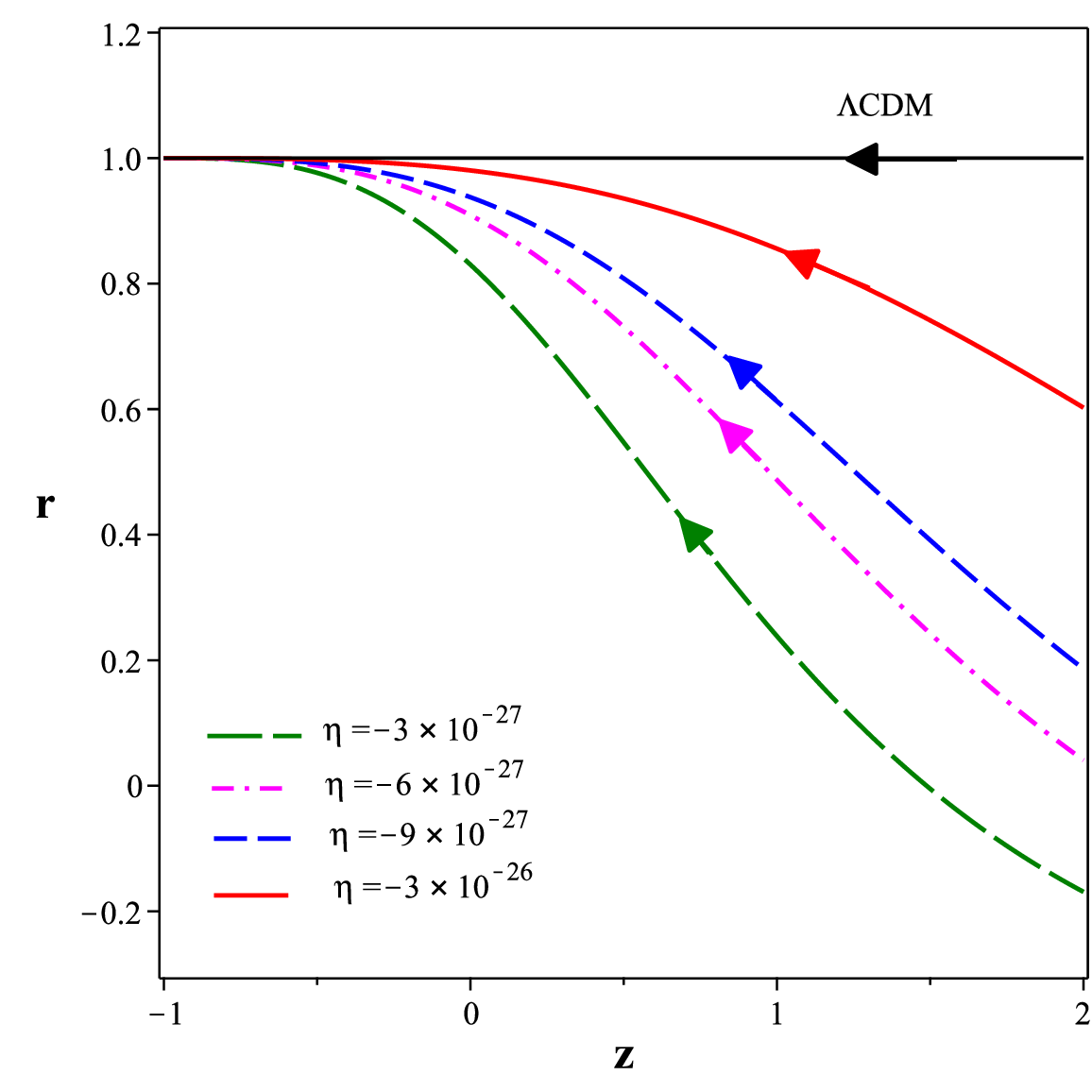}\includegraphics{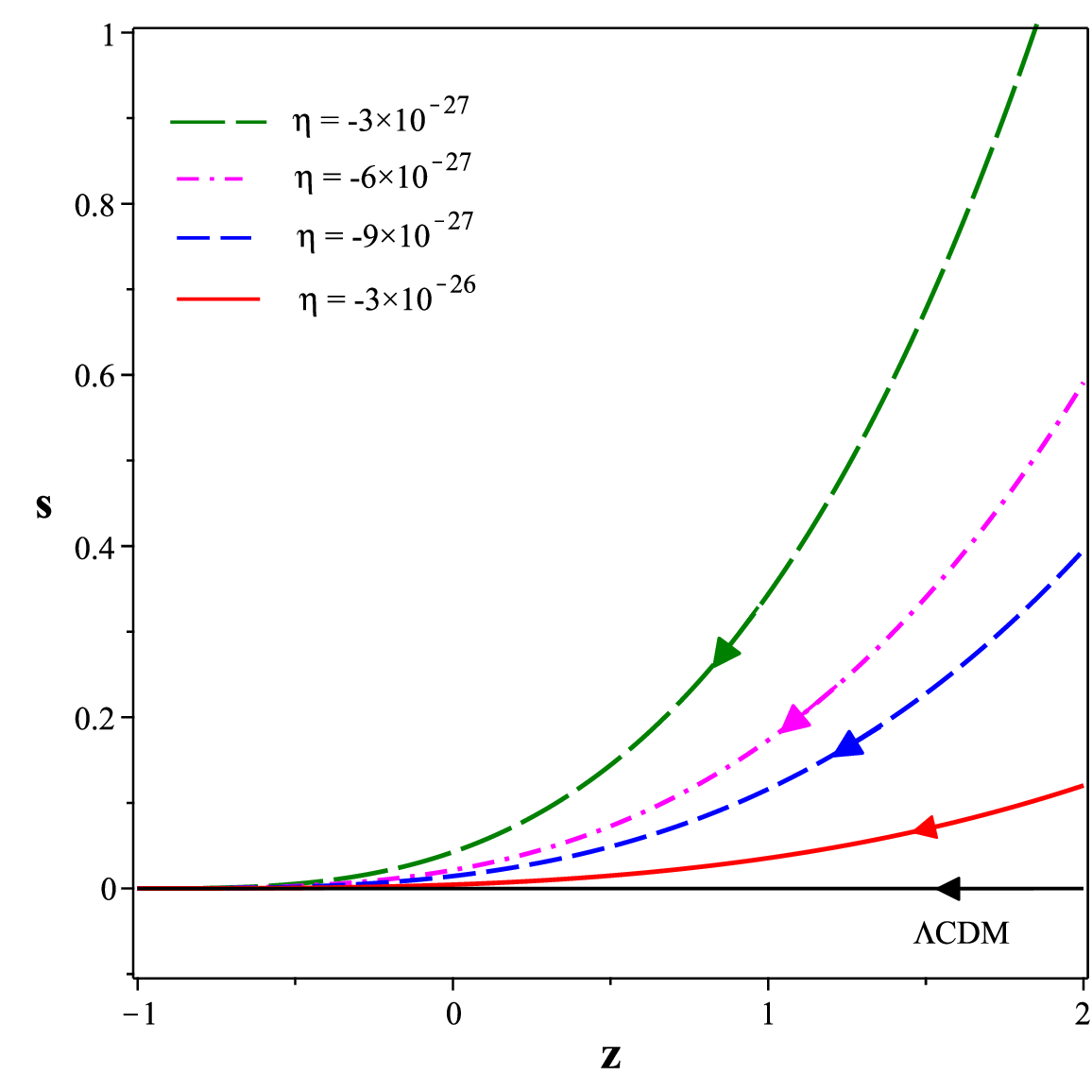}\vspace{6.5cm}
		\end{center}
		\caption{\label{fig7}\small {The evolutionary trajectories of $r$ versus $z$ (left panel) and
				$s$ versus $z$ (right panel) for different values of $\eta$.}}
	\end{figure}
	
	\newpage
	\section{\label{sec7}$O_{m}(z)$ Diagnostic}
	Another well-known diagnostic for the dark energy models is the $O_{m}$ diagnostic~\cite{Sahni2008}.
	This alternative route to statefinder diagnostics distinguishes the $\Lambda$CDM from other dark energy
	models without direct recourse to the equation of state of the cosmic fluid. In this diagnostic route, one defines
	\begin{equation}
		O_{m}(x)\equiv \frac{h^{2}(x)-1}{x^{3}-1}\,,
	\end{equation}
	where $x=1+z$ and $h(x)=\frac{H(x)}{H_{0}}$.
	According to the modified Friedman equation (\ref{eq32}) obtained for a spatially flat FRW metric,
	we define a dimensionless Hubble parameter as follows
	\begin{eqnarray}
		\label{eq45}\Bigg(\frac{H(z)}{H_0}\Bigg)^2=\Omega_{_{0m}}(1+z)^3+(1-\Omega_{_{0m}})f(z)\,,
	\end{eqnarray}
	where
	\begin{eqnarray}
		\label{eq46}f(z)=\exp\Bigg[{3\int^{z}_{0}\Big(\frac{1+w(z')}{1+z'}\Big)dz'}\Bigg]\,,
	\end{eqnarray}
	and $\Omega_{_{0m}}$ represents the matter density at the current epoch. It is obvious that with the given values of the
	density parameters, there is a rigorous constraint on $w(z)$. In our model, the EoS parameter $w(z)$, which contains
  quantum gravitational effect is specified in (\ref{eq39}). In our setup with quantum gravitational effect
	encoded in EUP (\ref{eq23}) with negative deformation parameter, we obtain $O_{m}(z)$ diagnostic as follows
	\begin{eqnarray}
		\label{eq48}O_{m}(z)=-\Bigg(\frac{1}{\left(1+z\right)^{3}-1}\Bigg)
		\Bigg(\frac{4\eta\left({\cal{W}}\!\left(-\frac{1}{16\eta}\exp{\{\frac{-2\pi\,l_{pl}^{2}\rho_0\left(1+z\right)^{3}
					+3\eta}{3\eta}\}}\right)-1\right)^{2}}{{\cal{W}}\!\left(-\frac{1}{16\eta}\exp{\{\frac{-2\pi\,l_{pl}^{2}\rho_0\left(1+z\right)^{3}
					+3\eta}{3\eta}\}}\right)\mathit{H_0}^2}+1\Bigg)\,.
	\end{eqnarray}
	We have shown the behavior of $O_{m}$ evolution versus redshift $z$ in Fig. 8.
	According to the figure, as the redshift increases, the $O_{m}(z)$ decreases.
	The figure shows also that the late time IR quantum corrected cosmic history is the same as the $\Lambda$CDM.
	We have adopted the diagnostic characteristics of $O_{m}$ in terms of $\Omega_{0m}$ from the recent results of Planck2018~\cite{Pl18a}.
	
	\begin{figure}
		\begin{center}\includegraphics{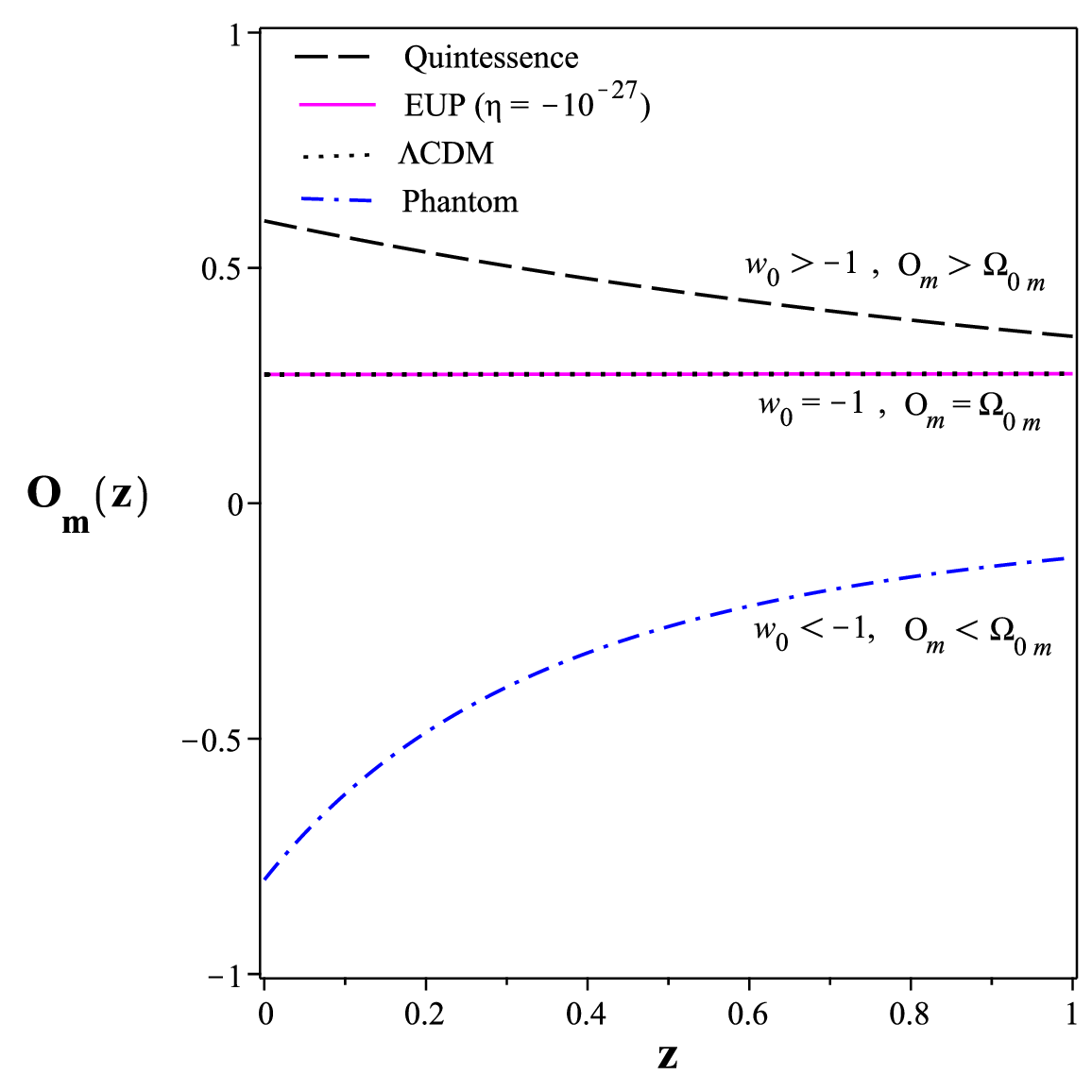}\vspace{6.5cm}
		\end{center}
		\caption{\label{fig8}\small {Behavior of the $O_{m}$ diagnostic versus $z$.}}
	\end{figure}
	\section{\label{sec8}Hubble function and infrared quantum gravitational effect}
	Since the Hubble parameter is deduced only from astrophysical observations and
	does not depend on any type of cosmological models in essence (however, remember the Hubble
	tension and related topics~\cite{Capozziello2020,Vagnozzi2023}), one of the most appropriate
	cosmological probe is to study the Hubble parameter as a function of redshift.
	The evolution of the universe in the presence of the infrared quantum
	gravitational effect gives the Hubble parameter $H(z)$ via Eq. (\ref{eq32}) that can be written as follows
	\begin{eqnarray}
		\label{eq49}H(z)=-2\eta\Bigg\{\frac{{\cal{W}}\!\left(-\frac{1}{16\eta}
			\exp\{-\frac{2\pi\,l_{pl}^{2}\rho_{0}(1+z)^3-3\eta}{3\eta}\}\right)-1}{\sqrt{\eta\,{\cal{W}}\!
				\left(-\frac{1}{16\eta}\exp\{{-\frac{2\pi\,l_{pl}^{2}\rho_{0}(1+z)^3-3\eta}{3\eta}}\}\right)}}\Bigg\}\,.
	\end{eqnarray}
	
	\begin{figure}
		\begin{center}\includegraphics{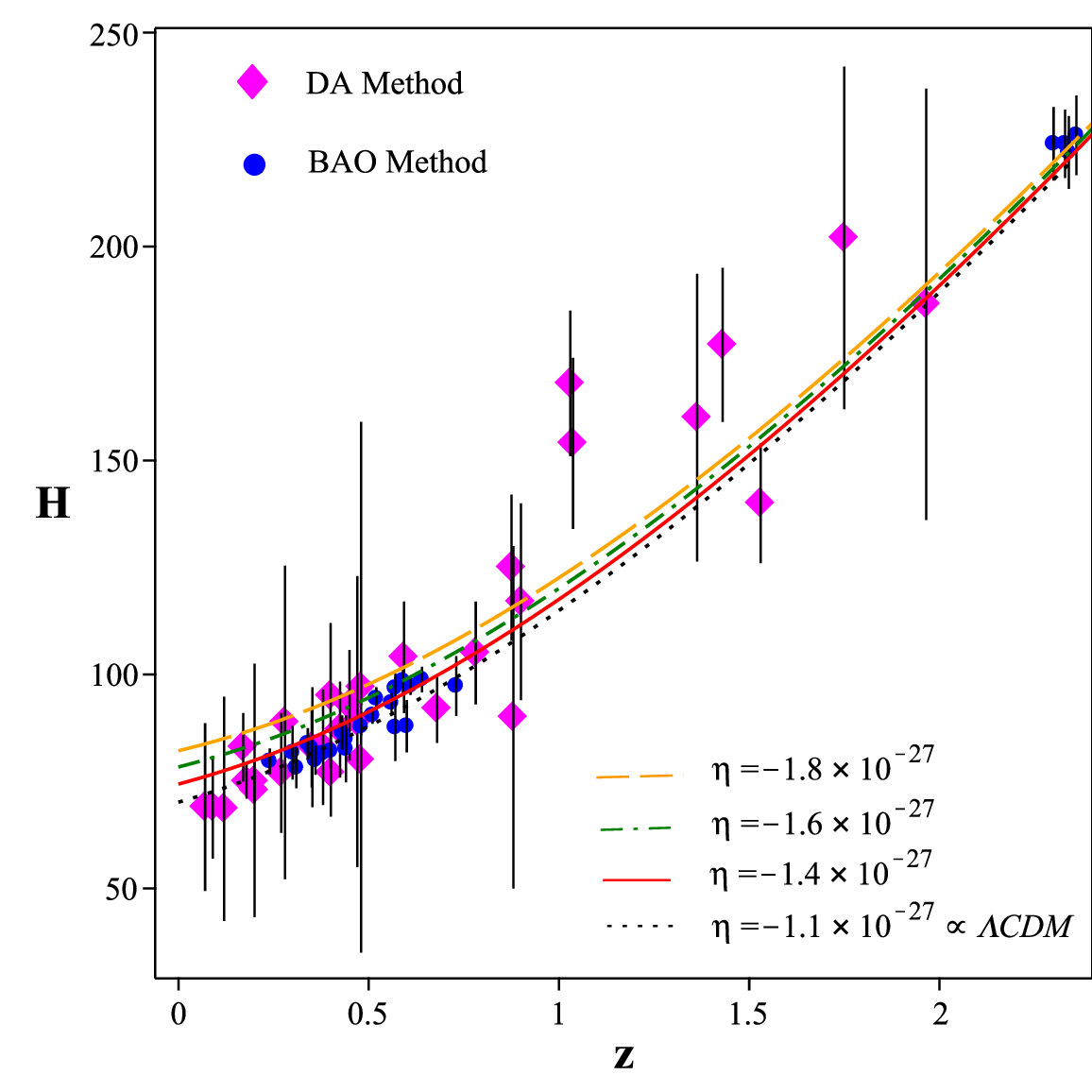}\vspace{8cm}
		\end{center}
		\caption{\label{fig9}\small {Evolution of the Hubble parameter with redshift for different values of $\eta$.
				Observational values of $H(z)$ are taken from~\cite{Kale2023}.}}
	\end{figure}
	Fig. 9 demonstrates the most appropriate curves of the Hubble's parameter in terms of redshift $z$ using $57$
	points of $H(z)$ data including $31 (DA) + 26 (BAO + \textrm{other})$ ~\cite{M.Koussour2022}. Also, we have compared
	our model with $\Lambda$CDM and as is evident, our model with infrared quantum gravitational effect is a successful alternative to $\Lambda$CDM model.
	In what follows, we use the cosmographic parameters in order to study the evolution of the universe via cosmological distance measures.
	
	\section{\label{sec9}Distance Measures}
	One of the fundamental challenges in cosmology is the precise measurement of cosmic distances.
	Distance measurements help to provide an almost comprehensible natural picture of the distance
	between two objects or two events in the universe. Here we study various measurement schemes
	of cosmic distances, taking into account the quantum gravitational effect at large distances which is finally an inevitable
	part of the puzzle of the ultimate theory of quantum gravity. We start with the cosmographic parameters formulas
	and then describe EUP-corrected distance measures in our setup.
	
	The ``Hubble distance" is defined by
	\begin{eqnarray}
		\label{eq50}d_{H}=\frac{c}{\mathit{H_0}}\,.
	\end{eqnarray}
	where, $H_0$ is the current Hubble constant which shows the constant of proportionality between speed
	of light $c$ and Hubble distance $d_H$ in the expanding universe.
	Based on the modified Friedmann equation, we determine a dimensionless Hubble parameter as
	\begin{eqnarray}
		\label{eq51}E_{z}=\frac{\mathit{H(z)}}{\mathit{H_0}}=\sqrt{\Omega_m(1+z)^3+\Omega_{EUP}(z)}\,,
	\end{eqnarray}
	where, $\Omega_m$ and $\Omega_{EUP}$ are normalized values of the current matter-energy density
	and an effective density parameter attributed to the EUP effect respectively as follows
	\begin{eqnarray}
		\label{eq52}\Omega_{m}=\frac{8\pi G}{3}\frac{\rho_{0}}{H_0^2}\,.
	\end{eqnarray}
	
	\begin{eqnarray}
		\label{eq53}\Omega_{EUP}(z)=\frac{8\pi G}{3}\frac{\rho_{_{EUP}}}{H_0^2}\,.
	\end{eqnarray}
	Note that, $\rho_{_{EUP}}$ which is given as Eq. (\ref{eq31}) is a function of the redshift.
	Therefore, the density parameter dependent on the EUP effect has a dynamic character. In the next stage,
	we explore the effect of the IR cutoff encoded in EUP on other distance measures.
	\subsection{Comoving distance}
	
	Comoving distance is implemented in cosmology to measure distances between cosmological objects
	and to explore the state of expansion of the universe. It is also applied to determine the volume
	of the observable universe. According to the dimensionless Hubble parameter Eq. (\ref{eq51})
	that we have introduced in the previous section, the comoving distance is specified as follows
	\begin{eqnarray}
		\label{eq54}
		d_{c}(z)=d_{H}\int_0^z\frac{dz'}{E(z')}\,,
	\end{eqnarray}
	where $d_{H}$ is the Hubble distance described by Eq. (\ref{eq50}).
	Now, considering the large distance quantum gravitational effect we modify the comoving distance as follows
	\begin{eqnarray}
		\label{eq55}
		d_{c}(z)=-\frac{\sqrt{-\eta{\cal{W}}(-\frac{0.17}{\eta})}\,
			\Bigg(12\eta\Big({\cal{W}}\!\,(-\frac{0.17}{\eta})-1\Big)
			+\pi\rho_0(z+2)(z^{2}+2z+2)l_{pl}^{2}\Bigg)\,c\,z}
		{24\,\eta^{2}\Big({\cal{W}}\!\,(-\frac{0.17}{\eta})-1\Big)^{2}}+O(l_{pl}^4)\,.
	\end{eqnarray}
	Note that, the comoving distance is a basic distance measure in cosmography since, as is evident in what follows,
	all other distances are straightforwardly deducible in terms of this measure.
	Figure 10 shows the comoving distance in comparison with other measure distances versus the redshift.

	\subsection{Transverse comoving distance}
	
	Transverse comoving distance $d_M$ is directly corresponded to the comoving distance as
	\begin{equation}
		\label{eq56}
		d_M(z)=\left\{\begin{array}{ll}\frac{d_H}{\sqrt{\Omega_k}}\sinh(\frac{\sqrt{\Omega_k}d_c(z)}{d_H})\quad \quad \Omega_k>0 \\  \\
			d_c(z) \quad \quad \quad\quad\quad\quad\quad\quad\Omega_k= 0 \\ \\ \frac{d_H}{\sqrt{|\Omega_k|}}\sinh(\frac{\sqrt{|\Omega_k|}d_c(z)}{d_H})\quad \Omega_k<0
		\end{array}\right.
	\end{equation}
	Since in our setup we assumed $\Omega_k=0$, therefore we have $d_M(z)=d_{c}(z)$.
	Figure 10 shows the transverse comoving distance in comparison with other measure distances versus the redshift.
	
	\subsection{Angular diameter distance}
	
	The angular diameter distance to an object at redshift $z$ is given by
	\begin{eqnarray}
		\label{eq57}d_A(z)=\frac{d_{M}(z)}{1+z}\,,
	\end{eqnarray}
	where $d_{M}(z)$ is given by Eq. (\ref{eq56}). Therefore, we have
	
	\begin{eqnarray}
		\label{eq58}d_A(z)=-\frac{\sqrt{-\eta{\cal{W}}(-\frac{0.17}{\eta})}\,
			\Bigg(12\eta\Big({\cal{W}}\!\,(-\frac{0.17}{\eta})-1\Big)
			+\pi\rho_0(z+2)(z^{2}+2z+2)l_{pl}^{2}\Bigg)\,c\,z}
		{24\,\eta^{2}\Big({\cal{W}}\!\,(-\frac{0.17}{\eta})-1\Big)^{2}(1+z)}+O(l_{pl}^4)\,.
	\end{eqnarray}
	Figure 10 shows the angular diameter distance in comparison with other measure distances versus the redshift.
	
	\begin{figure}
		\begin{center}\includegraphics{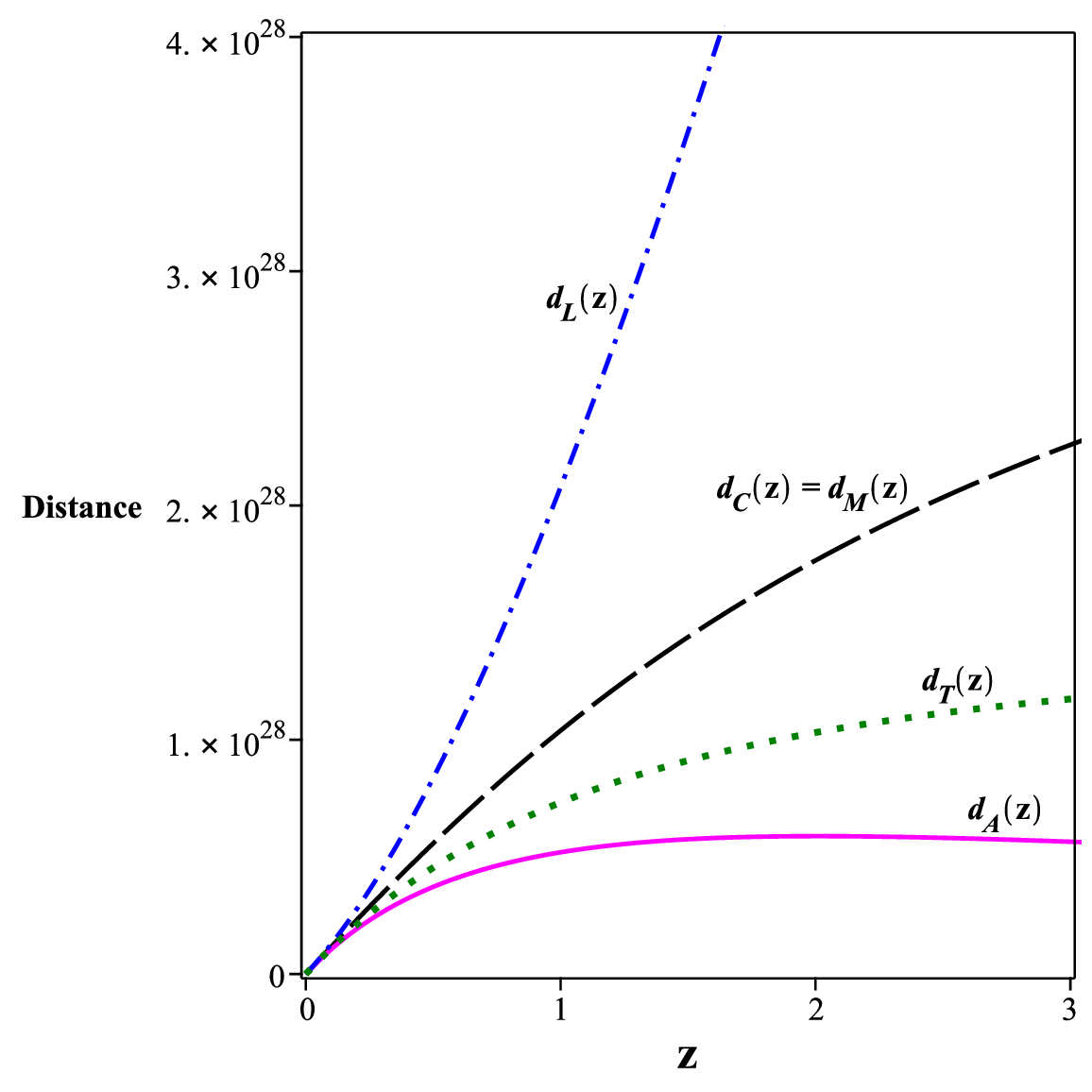}\vspace{6.5cm}
		\end{center}
		\caption{\label{fig10}\small {A comparison of cosmological distance measures from $z=0$ to $z=3$ for $\eta=-3\times10^{-27}$.}}
	\end{figure}
	
	\subsection{Luminosity Distance}
	
	The luminosity distance is a measure of distance that shows how far away an astronomical object
	is predicated on its luminousness and visible brightness. It is distinct from other distance
	measurements, including comoving distance and angular diameter distance, because it takes
	into consideration the expansion of the universe. The luminosity distance $d_L$ in terms of the redshift
	appropriate to the transverse comoving distance and angular diameter distance is given by ~\cite{weinberg1972,weedman1986}
	\begin{eqnarray}
		\label{eq59}d_L(z)=(1+z)d_M(z)=(1+z)^2d_{A}(z)\,,
	\end{eqnarray}
	The luminosity distance in the presence of the large distance quantum gravitational effect encoded in EUP is as follows
	\begin{eqnarray}
		\label{eq60}d_L(z)=-\frac{\sqrt{-\eta{\cal{W}}(-\frac{0.17}{\eta})}\,
			\Bigg(12\eta\Big({\cal{W}}\!\,(-\frac{0.17}{\eta})-1\Big)
			+\pi\rho_0\,l_{pl}^{2}(z+2)(z^{2}+2z+2)\Bigg)\,c\,z(1+z)}
		{24\,\eta^{2}\Big({\cal{W}}\!\,(-\frac{0.17}{\eta})-1\Big)^{2}}+O(l_{pl}^4)\,.
	\end{eqnarray}
	Fig. 11 shows the plot of the EUP-corrected luminosity distance as a function of the
	redshift based on Eq. (\ref{eq60}) for different values of the EUP parameter, $\eta$.
	Increasing the parameter $\eta$ increases the luminosity distance (note that $\eta$ is a negative quantity).
	\begin{figure}
		\begin{center}\includegraphics{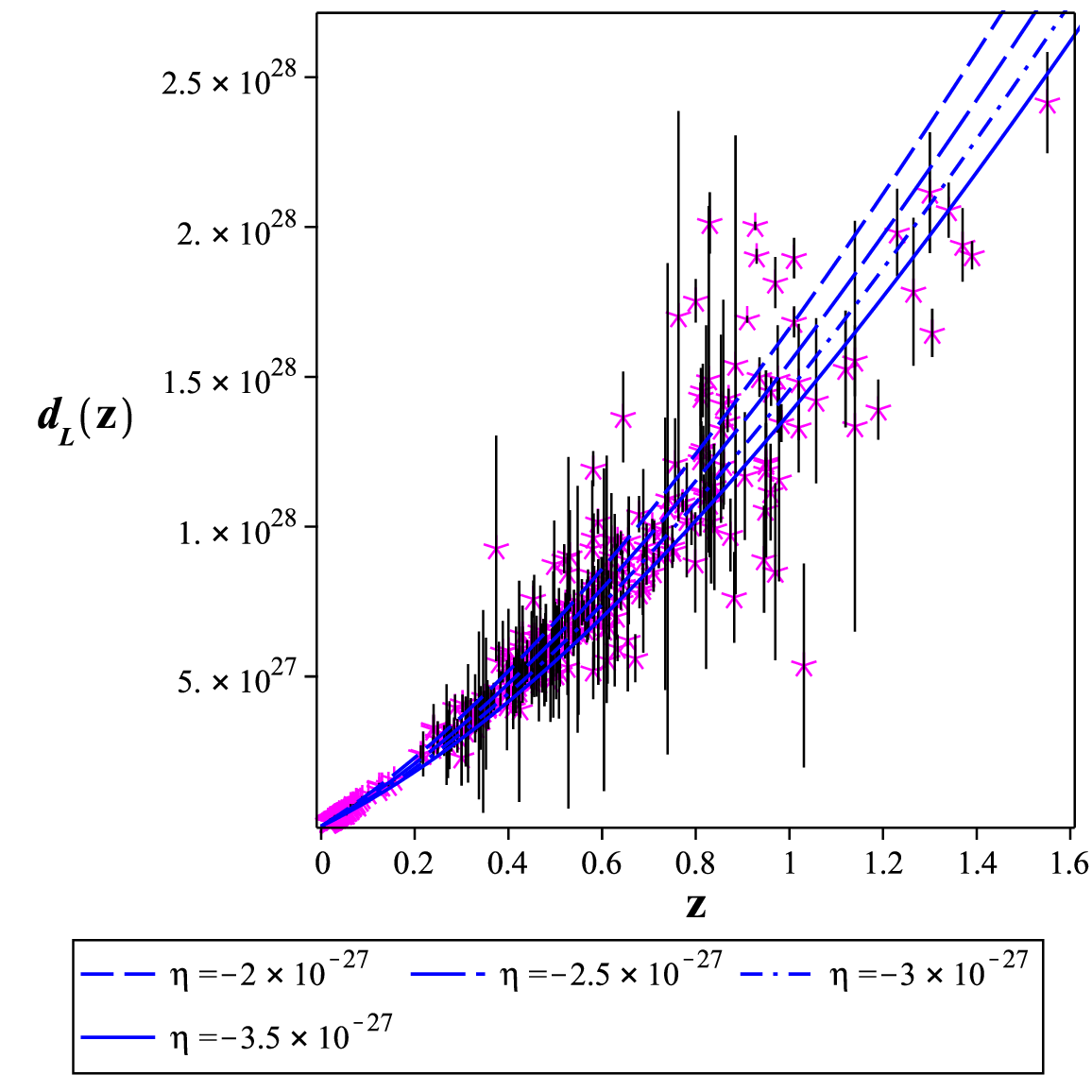}\vspace{6.5cm}
		\end{center}
		\caption{\label{fig11}\small {Luminosity distance as a function of the redshift for different values of $\eta$.
				The used sample includes 397 supernovae. Luminosity distances and errorbars were calculated with $M_{B}=-19.44$,
				$\alpha=0.104^{+0.018}_{-0.018}$ and $\beta=2.48^{+0.1}_{-0.12}$ from Ref.~\cite{M.Hicken2009}. }}
	\end{figure}
	
	\subsection{Light-travel distance}
	
	The light-travel distance measure which is closely related to other distance measures, is defined as follows
	
	\begin{eqnarray}
		\label{eq61}d_T(z)=dH\int^{z}_{0}\frac{dz'}{(1+z')E(z')}\,,
	\end{eqnarray}
	where in the presence of large distance quantum gravity effect encoded in EUP takes the following form
	\begin{eqnarray}
		\label{eq62}d_T(z)=-\frac{\sqrt{-\eta{\cal{W}}(-\frac{0.17}{\eta})}\Bigg(3\eta\Big({\cal{W}}\!
			\,(-\frac{0.17}{\eta})-1\Big)
			+\pi\,\rho_0\,l_{pl}^{2}\,(1+z)^3\Bigg)\,c\,z}{6\,\eta^{2}(1+z)
			\Big({\cal{W}}\!\,(-\frac{0.17}{\eta})-1\Big)^{2}}+O(l_{pl}^4)\,.
	\end{eqnarray}
	Figure 10 shows the light-travel distance in comparison with other measure distances versus the redshift.
	
	\section{\label{sec9}Distance Modulus}
	One of the well-known methods used to represent distances in astronomy is the distance module.
	It characterizes distances on a logarithmic scale established on the astronomical magnitude system.
	Distance modulus $\mu$ of a system is described as
	\begin{eqnarray}
		\label{eq63}\mu(z)=5 \log{\Big(d_L(z)\Big)} +\mu_0\,,
	\end{eqnarray}
	where, $\mu_{0}=5 \log{\Big(\frac{{H_0^{-1}}}{1Mpc}\Big)}+25$.
	Using Eq. (\ref{eq60}), we obtain the following expression for the distance modulus
	\begin{eqnarray}
		\label{eq64}\mu(z)=\frac{5\ln\!\left(-\frac{\sqrt{-\eta W\left(-\frac{{0.17}}{\eta}\right)}\,\left(12\eta \left(W\left(-\frac{{0.17}}{ \eta}\right)-1\right)+l^{2}\pi \left(z +2\right)\left(z^{2}+2z+2\right)\rho_0\right)c\left(1+z\right)z}{24\mathit{H_0}\mathit{Mpc}\,\eta^{2} \left(W\left(-\frac{{0.17}}{\eta}\right)-1\right)^{2}}\right)}{\ln\!\left(10\right)}+25\,.
	\end{eqnarray}
	The authors in Ref.~\cite{M.Hicken2009} have used the following equation to obtain the SALT
	output parameters and calculated the most suitable form
	\begin{eqnarray}
		\label{eq62}\mu(z)=\mu^{max}_{B}-M_{B}+\alpha(s-1)-\beta C\,,
	\end{eqnarray}
	where, $m_{B}$ is magnitude in $B$ band, $M_{B}$ is marginalized,
	$\alpha$ and $\beta$ are empirical coefficients, $s$ is a time stretch
	factor and $C$ is a color parameter. To obtain the distances to each SNIa,
	the authors use the 1-d marginalization coefficients, $\alpha$, $\beta$ , and $M$.
	Figure 12 shows the distance modulus versus the redshift for different values of $\eta$ in our setup.
	
	\begin{figure}
		\begin{center}\includegraphics{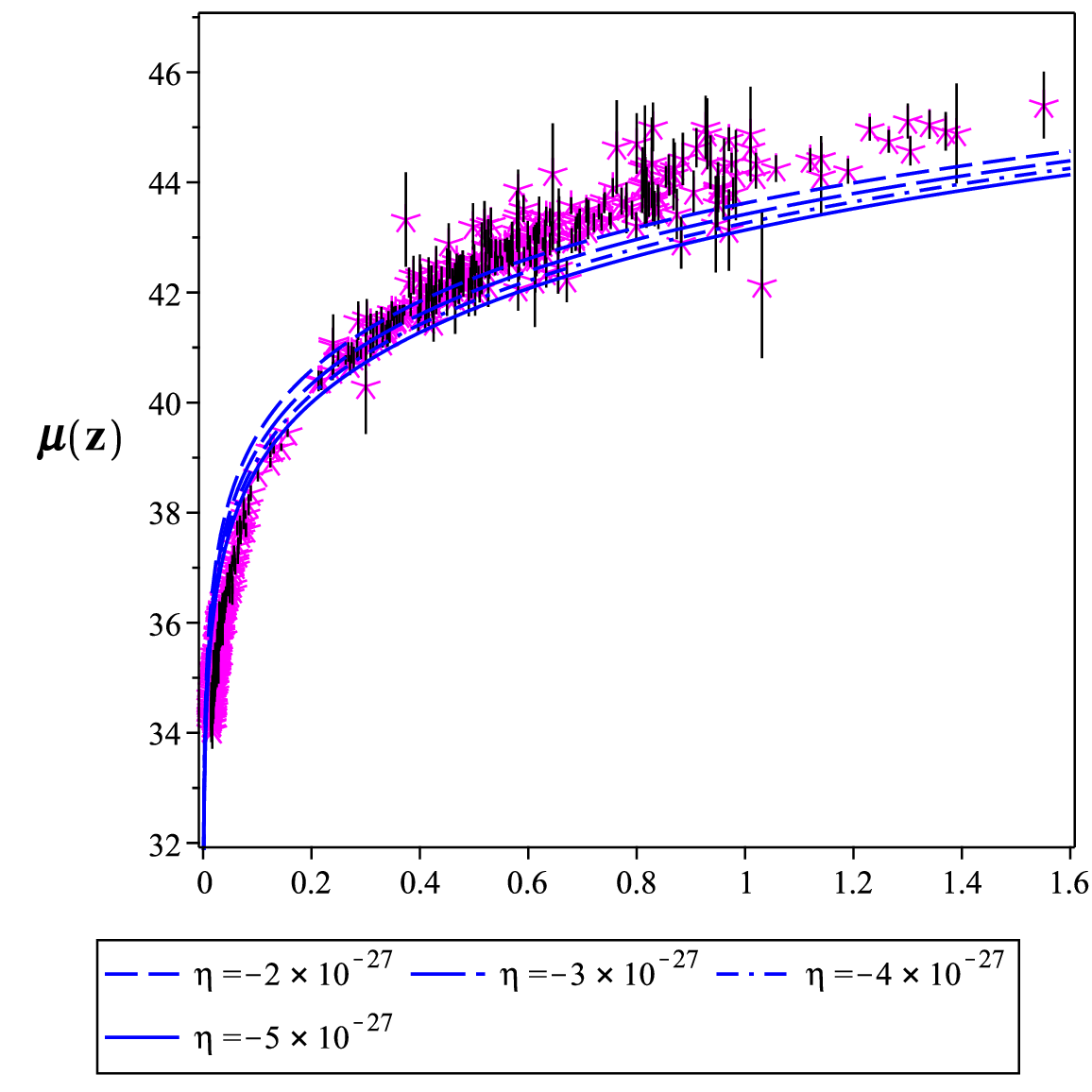}\includegraphics{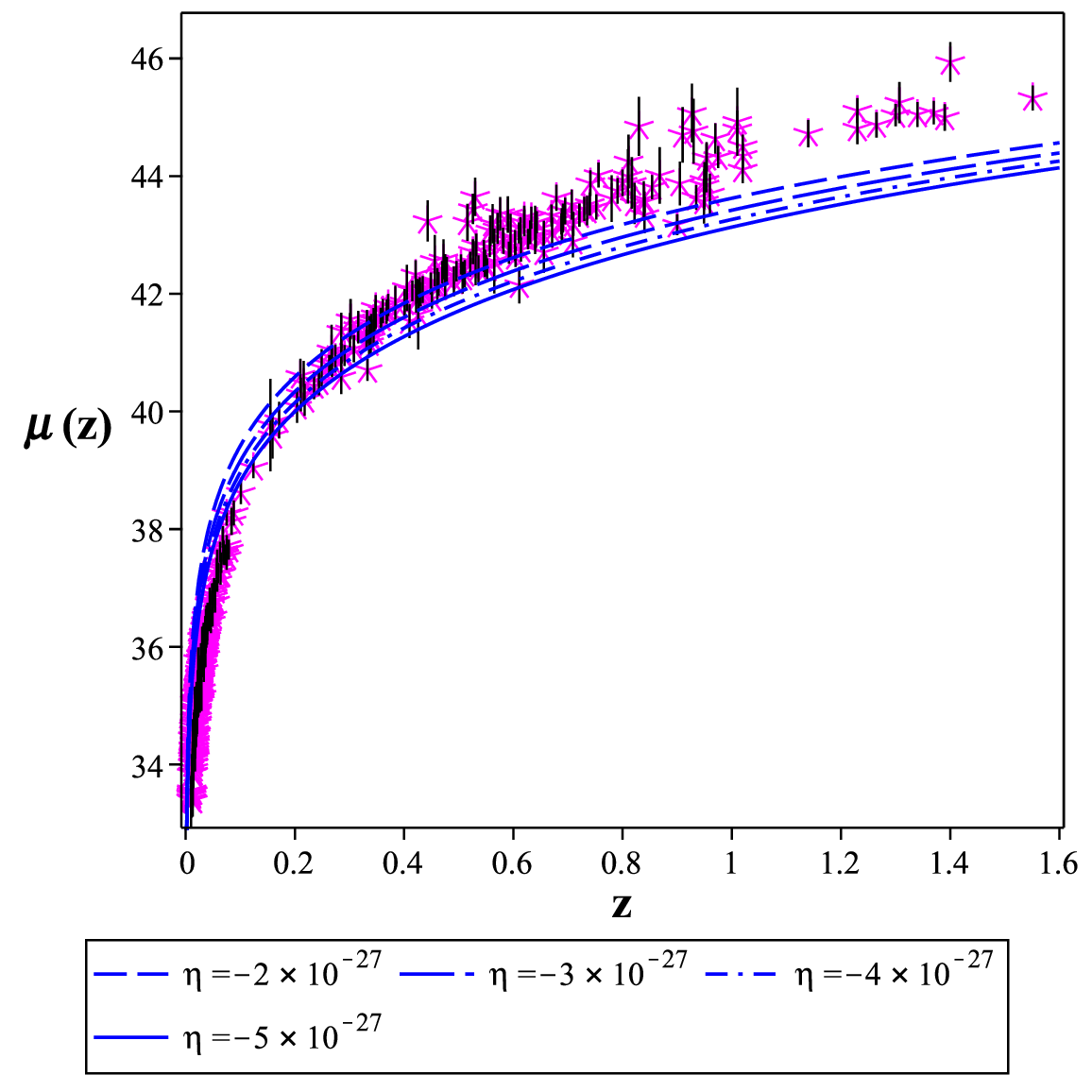}\vspace{6.5cm}
		\end{center}
		\caption{\label{fig12}\small {Distance modulus as a function of redshift for different values of $\eta$.
				These distance modulus were calculated with $M_{B}=-19.46$, $\alpha=0.34^{+0.08}_{-0.08}$ and
				$\beta=2.59^{+0.12}_{-0.08}$ for 397 points of supernova data (the left panel) and with $M_{B}=-19.44$,
				$\alpha=0.104^{+0.018}_{-0.018}$ and $\beta=2.48^{+0.1}_{-0.12}$ for 351 points of supernovae (the right panel) following Ref.~\cite{M.Hicken2009}.}}
	\end{figure}
	Finally, figure 13 illustrates the distance modulus versus the redshift with some other probes (following Ref.~\cite{Gupta2019}) that specifies explicitly the $\Lambda$CDM model.
	\begin{figure}
		\begin{center}\includegraphics{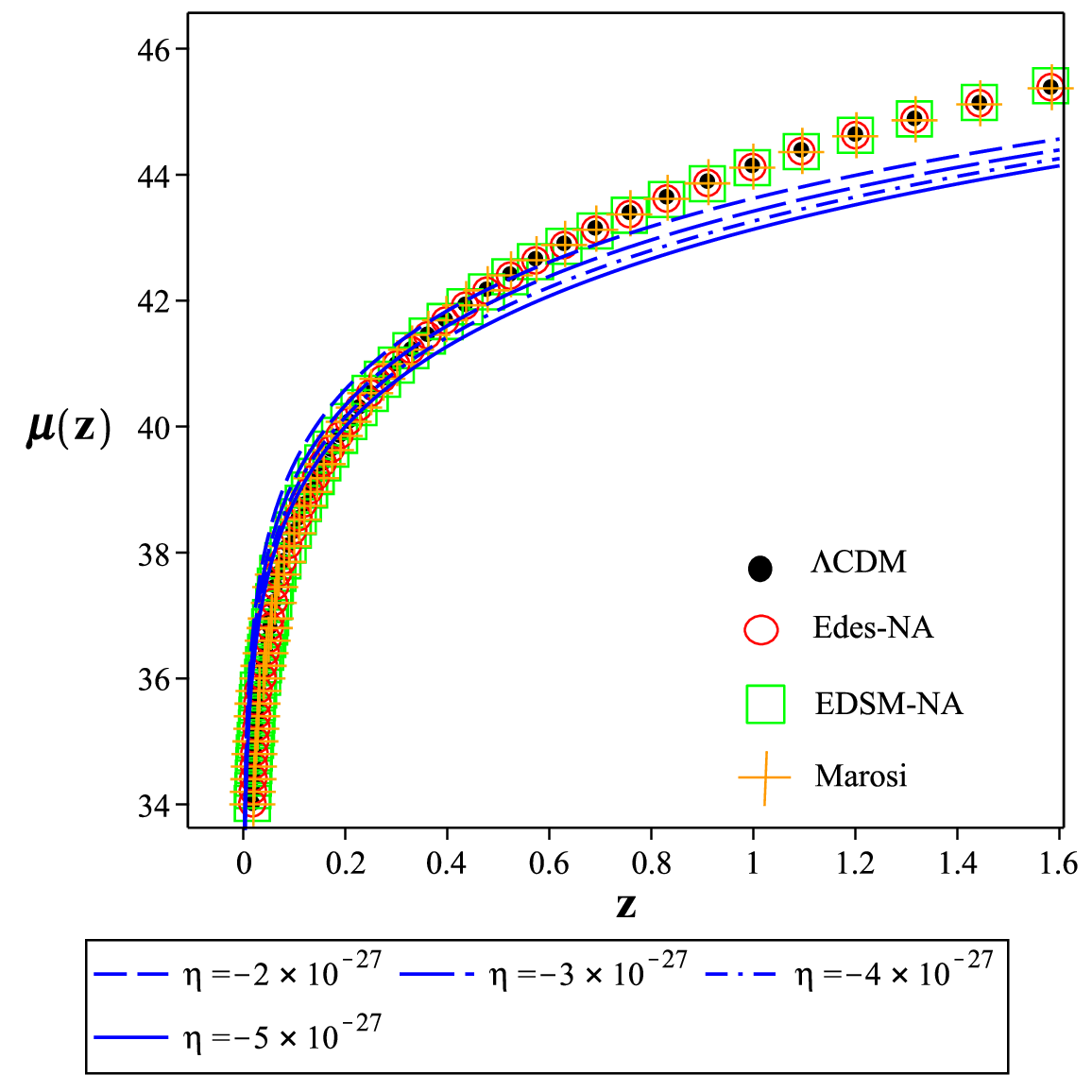}\vspace{6.5cm}
		\end{center}
		\caption{\label{fig13}\small {Illustration of the distance modulus versus the redshift with some other probes (following Ref.~\cite{Gupta2019}). The $\Lambda$CDM model is specified explicitly.}}
	\end{figure}

	\section{\label{11}Summary}

	We studied the impact of large distance quantum gravity effect on the late time cosmological dynamics via an
	extended uncertainty principle with a negative deformation parameter. To do so, we considered the first law of thermodynamics for the
	apparent horizon of a spatially flat Friedman-Robertson-Walker universe inspired by the famous black hole
	entropy-area relation. We derived the EUP-corrected Friedmann equations through the mentioned thermodynamic
	approach. We show that these EUP-modified cosmological equations have very distinct impacts on the late time cosmic evolution.
	To explore these impacts, we focused on the evolution of cosmological observables and also distance measures in
	this setup. A close inspection of the Hubble rate, deceleration parameter, and effective equation of state parameter
	reveals the late time cosmic speed up and transition to a phantom phase in this setup with just ordinary, non-relativistic matter.
	So, EUP modification with a negative deformation parameter suffices to explain late time cosmic speed-up without recourse to dark energy.
	To deepen our investigation, we also focused on the impacts of this EUP on the cosmological distance measures.
	By using the existing observational data sources, we demonstrated how this EUP-corrected scenario is feasible from the
	observational viewpoint. By treating the constructed setup in a diagnostic playground (via statefinder and $O_m$
	diagnostics), we proved the possibility of realization of the $\Lambda$CDM model as the concordance model in this
	EUP-corrected scenario. To summarize, in this paper we proved that a model universe containing non-relativistic standard matter as a dominant
	ingredient with quantum gravity effect encoded in extended uncertainty principle with negative deformation parameter is a suitable model
     universe to mimic the late time universe and this is supported by the most recent observational data.\\\\
	
	{\bf Acknowledgement}\\
    We appreciate the referee for insightful comments and helpful suggestions. We alse appreciate Subir Ghosh for constructive discussion.

\end{document}